\documentclass[12pt, preprint]{aastex}

\begin{document}

\def\be{\begin{equation}}
\def\ee{\end{equation}}
\def\lesssim{\raisebox{-0.3ex}{\mbox{$\,\, \stackrel{<}{_\sim} \,\,$}}}
\def\gtrsim{\raisebox{-0.3ex}{\mbox{$\stackrel{>}{_\sim} \,$}}}
\def\EB{\hbox{${\rm {\bf E} \times {\bf B}}$}}
\def\( {\left( }
\def\) {\right) }

\title{Formation of a Partially-Screened Inner Acceleration Region in Radio Pulsars: Drifting Subpulses and Thermal X-Ray Emission from Polar Cap Surface}
\author{Janusz Gil\altaffilmark{1,2},
George Melikidze\altaffilmark{1,3}
\& Bing Zhang
\altaffilmark{2}}
\altaffiltext{1}{Institute of Astronomy, University of Zielona
G\'ora, Lubuska 2, 65-265, Zielona G\'ora, Poland}
\altaffiltext{2}{Department of Physics, University of Nevada, Las
Vegas, USA} \altaffiltext{3}{CPA, Abastumani Astrophysical
Observatory, Al. Kazbegi ave. 2a, 0160, Tbilisi, Georgia}

\begin{abstract}
The subpulse drifting phenomenon in pulsar radio emission is considered within the partially screened inner gap model,
in which the sub-Goldreich-Julian thermionic flow of iron ions or electrons coexists with the spark-associated
electron-positron plasma flow. We derive a simple formula that relates the thermal X-ray luminosity $L_{\rm x}$ from
the spark-heated polar cap and the \EB\ subpulse periodicity $\hat{P}_3$ (polar cap carousel time). For PSRs B0943+10
and B1133+16, the only two pulsars for which both $\hat{P}_3$ and $L_{\rm x}$ are known observationally, this formula
holds well. For a few other pulsars, for which only one quantity is measured observationally, we predict the value of
the other quantity and propose relevant observations that can confirm or discard the model. Then we further study the
detailed physical conditions that allow such partially screened inner gap to form. By means of the condition $T_{\rm
c}/T_{\rm s}>1$ (where $T_{\rm c}$ is the critical temperature above which the surface delivers a thermal flow to
adequately supply the corotation charge density, and $T_{\rm s}$ is the actual surface temperature), it is found that a
partially-screened gap (PSG) can be formed given that the near surface magnetic fields are very strong and curved. We
consider both curvature radiation (CR) and resonant inverse Compton scattering (ICS) to produce seed photons for pair
production, and find that the former is the main agency to produce gamma-rays to discharge PSG.
\end{abstract}


\section{Introduction}

Pulsar radio emission typically occurs in a form of periodic series of narrow bursts of radiation. These burst often
have a complex structure, in which higher order periodicities can be found. The individual pulses consist of one, few
to several subpulses. In some pulsars the subpulses demonstrate a very systematic drift across the pulse window. If the
pulses are folded with the basic pulsar period then the drifting subpulses form amazingly spectacular patterns called
drift-bands.

The phenomenon of drifting subpulses is a long standing puzzle in the pulsar research and its solution would likely
result in deeper understanding to the nature of pulsar radiation. It is generally believed that this phenomenon is
inherently associated with the so-called inner acceleration region above the polar cap, in which the magnetospheres
plasma does not corotate with the neutron star surface. The first model based on this idea was proposed by \citet[][
RS75 henceforth]{rs75}. The predictions of RS75 model were successfully compared with a handful of pulsars known to
show this phenomenon at that time. A decade later, \citet[][ R86 hereafter]{r86} compiled a list of about 40 drifting
pulsars, but drifting subpulses were still regarded as some kind of exceptional phenomenon. However, recently \citet[][
WES06 hereafter]{wes06} presented the results of a systematic, unbiased search for subpulse modulation in 187 pulsars
and found that the fraction of pulsars showing drifting subpulse phenomenon is likely to be larger than 55\%. They
identified 102 pulsars with drifting subpulses in their sample, with a large fraction of newly discovered drifters. The
authors concluded that the conditions required for the drifting mechanism to work cannot be very different from the
emission mechanism of radio pulsars. WES06 then suggest that the subpulse drifting phenomenon is an intrinsic property
of the pulsar emission mechanism, although drifting could in some cases be very difficult or even impossible to detect
due to insufficient signal-to-noise ratio. It is therefore essential to attempt to unravel the physical conditions that
can lead to formation of an inner acceleration region above the polar cap that could lead to development of the
subpulse drift phenomenon.

The classical vacuum gap model of RS75, in which spark-associated sub-beams of subpulse emission circulate around the
magnetic axis due to $\mathbf{E}\times\mathbf{B}$ drift of spark plasma filaments, provides the most natural and
plausible explanation of drifting subpulse phenomenon. However, despite its popularity, it suffers from the so-called
binding energy problem. Namely, under canonical conditions the surface charges (ions or electrons) are likely to be
directly pulled out of the surface so that a pure vacuum gap is difficult to form. The alternative steady flow polar
cap models, the so-called space-charge-limited flow models \citep{as79,hm98}, cannot give rise to the intermittent
``sparking'' behavior, which seems necessary to explain the subpulse drift phenomenon in radio pulsars. \citet[][ GM01
hereafter]{gm01} revisited the binding energy problem of RS75 model and argued that the formation of the vacuum gap
(VG) is, in principle possible, although it requires a very strong non-dipolar surface magnetic fields, much stronger
than a canonical dipolar component inferred from the observed spindown rate. Once the binding is strong enough to
prevent the thermionic emission at the full space charge limited flow, the inevitable \EB\ drift of plasma filaments
will result in the observable subpulse drift phenomenon. It has been known for a long time that in order to allow all
radio pulsars to produce electron-positron pairs (the necessary condition for coherent radio emission), the
near-surface magnetic fields must include multipole components dominating over global dipole field
\citep{rs75,as79,zhm00}. Growing observational evidence of non-dipolar structure of surface magnetic field
\footnote{It is worth mentioning that although RS75 implicitly assumed non-dipolar surface magnetic fields to treat the $\gamma$-$B$
pair production processes, in their calculations of many other physical quantities (such as the surface charge density)
they still used dipolar form, presumably for the sake of simplicity.} accumulates, and the suggestion that such a
sunspot-like fields form during the early proto-neutron star stage has been proposed
\footnote{Another possible mechanism of creating small scale anomalies of surface magnetic fields was proposed by \citet{grg03}.
They argued that due to a Hall drift instability, the poloidal magnetic field structures can be generated from
strong subsurface toroidal fields.}
\citep[e.g.][]{ug04}. \citet[][ GM02 hereafter]{gm02} calculated the non-dipolar VG
model for 42 drifting subpulse pulsars tabulated by R86 and argued that VG can be formed in all considered pulsars,
provided that the actual surface magnetic field was close to $10^{13}$ G independently of the value of the canonical
dipolar magnetic field.

Although the binding energy problem could be, at least in principle, resolved by assuming an appropriately strong
surface magnetic field, yet another difficulty of the RS75 model was that it predicted a much too fast \EB\ drift rate.
Motivated by this issue, \citet[][ GMG03 hereafter]{gmg03} developed further the idea of the inner acceleration region
above the polar cap by including the partial screening by a sub-Goldreich-Julian thermal flow from the surface due to
the spark-associated polar cap heating. This idea was first introduced by \citep[ CR80 henceforth]{cr80}, who argued
that even with thermionic ions included in the flow, the condition above the polar cap is close to that of a pure
vacuum gap. A similar model was also invoked by \citet{um95,um96}. GMG03 reanalyzed this model and argued that a
thermostatic self-regulation should keep the surface temperature just few percent below the critical ion temperature at
which the gap potential drop is completely screened. This results in more than 90 \% of screening due to thermionic
emission. Given a similar gap height, the actual potential drop, and hence the \EB\ drift rate, is about 10 \% of that
of the pure vacuum case. This is still above the threshold for the magnetic pair production. Similar to the avalanche
pair production cascade introduced by RS75, the discharge of the growing potential drop above the polar cap would also
occur in the form of a number of sparks. In this paper we call such an inner accelerator a partially screened gap (PSG
henceforth). The latest XMM-Newton observation of the drifting pulsar PSR B0943+10 \citep[][ ZSP05 hereafter]{zsp05}
reveals a possible hot spot with the surface area much smaller than the conventional polar cap. This is consistent with
the polar cap heating from such a PSG, which at the same time gives the right \EB\ drift rate. This lends strong
support to the PSG model.

In this paper we study PSG model in greater detail and explore the physical conditions for the model to work. We also
apply this model to a new set of 102 pulsars from WES05 and show that it can work in every case, provided that the
surface non-dipolar magnetic field is strong enough, even stronger than $10^{13}$ G suggested by GM02 and GMG03. Our
new treatment is a combination of those used in GM02 and GMG03. Our working hypothesis is that drifting subpulses
manifest the existence of a thin inner acceleration region, with an acceleration length scale much shorter than the
polar cap size. The ultra-high accelerating potential drop discharges via a number of localized
$\mathbf{E}\times\mathbf{B}$ drifting sparks, as has been envisioned by RS75 . These sparks produce isolated columns of
electron-positron plasma that stream into the magnetosphere to generate radio-beams of the observed subpulse emission
due to some kind of plasma instability (see the \S8 for more discussion). Due to charge depletion with respect to the
co-rotational Goldreich-Julian (1969) value, sparks experience an unavoidable \EB\ drift with respect to the polar cap
surface. As a consequence, the spark-associated sub-beams of radio emission perform a slow circumferential motion that
is responsible for the observed subpulse drift. This model, which is often called ``a pulsar carousel model''
\citep{dr99}, is examined in this paper
\footnote{Other suggestions of subpulse drifting as phenomena occurring outside
the inner gap have been made \citep{kmm91,sa02,w03,gmm05,fkk06}, but the connection between the radio drifting rate and
the X-ray properties in those models is not yet clear, and we do not discuss them in the current paper.}. We are
particularly interested in the thermal effect associated with the surface bombardment by back-flowing particles
produced by sparks. The intrinsic drift rate (manifested by the tertriary drift periodicity) and the polar cap heating
rate (manifested by the thermal X-ray luminosity) should be correlated with each other, since they are determined by
the same value of the accelerating potential drop. The properties of drifting subpulses are discussed in \S2 and the
properties of charge depleted acceleration region above the polar cap are discussed \S3. We find a specific
relationship between the appropriate observables and conclud that it holds for a number of pulsars for which good
quality data is available (especially PSR B0943+10, ZSP05). It turns out that this relationship depends only on the
observational quantities and thus it is a powerful tool for testing this theoretical model. Although the number of
pulsars that have all necessary data for such testing is small at the moment, the clean prediction holds the promise to
ultimately confirm (or discard) the PSG model in the future. In \S\S4-7 we analyze this model in a more detailed manner
to investigate the microscopic conditions (e.g. near-surface configuration, radiation mechanism, etc) that are needed
to form such PSGs. The paper is summarized in \S8.

\section{Drifting subpulse phenomenon}

Let us briefly review the characteristic properties of the drifting subpulse phenomenon, in which subpulses typically
change phase from one pulse to another in a very organized manner, to the extent that they form apparent driftbands of
duration from several to a few tens of consecutive pulses. Usually more than one drift band appears, and the separation
between them measured in pulsar periods ranges from about 1 to about 15 \citep{b76,r86}. The subpulse intensity is
systematically modulated along drift bands, decreasing towards the edges of the pulse window. In some pulsars, however,
only periodic intensity modulations are observed, without any systematic phase change. These pulsars were identified as
those in which the line-of-sight cuts through a beam centrally \citep{b76}, thus showing a steep gradient of the
polarization angle curve \citep[e.g.][]{lm88}. On the other hand, the clear subpulse driftbands are typically found in
pulsars associated with the line-of-sight grazing the beam, thus showing a relatively flat position angle curve
\citep{b76,r86,lm88}. It is then obvious that both these kinds of subpulse behavior, i.e. systematic drift in phase and
phase stationary intensity modulation, represent the same phenomenon, namely, rotation of beams of subpulse emission
around the pulsar magnetic axis. The picture of rotating subbeams is also supported by spectral properties of the
observed radiation. The separation between drifting subpulses in a given pulse show dependence on frequency similar to
those of profile components and/or overall pulse width, while the observed periodicities related to patterns of
drifting subpulses are independent of radio frequency. This seems to exclude all frequency-dependent plasma effects as
a plausible source of the drifting subpulse phenomena. Also, a similarity of drift patterns at different radio
frequencies strongly suggest that the radiation of drifting subpulses is related to a relatively stable system of
isolated plasma filaments originating near the polar cap.

The observational characteristics of drifting subpulses described briefly above suggest then strongly an interpretation
of this phenomenon as a number of isolated sub-beams of radio emission, spaced more or less uniformly in the magnetic
azimuth, and rotating slowly around the magnetic axis. In terms of RS75 model this means that the subpulse-associated
sparks perform a circumferential motion around the pulsar beam axis. The most spectacular confirmation of this
"carousel model" was presented by \citet{dr99,dr01} and \citet{ad05}, who performed a sophisticated fluctuation spectra
analysis of single pulse data from PSR B0943+10 and detected clear spectral features corresponding to the rotational
behavior of subpulse beams. These results indicate clearly that sparks do not vanish after passing through the pulse
window in the form of drifting subpulses, but they continue to drift around the polar cap and reappear in the same
phase of the pulse window after a definite and measurable time period. The rotational features appearing in the
fluctuation spectra of PSR B0943+10 have extremely high values of $Q\sim 500$ (defined as the central frequency divided
by the width) confirming a circulational nature of the phenomenon. Similar feature with $Q\sim 100$ was found by
\citet{ad05} in PSR B0834+06. Quite recently \citet{ggk04} revealed a system of subpulse beams circulating around the
polar cap in PSR B0826-34. The high values of $Q$ suggests strongly that the trajectories of the spark motion
associated with drifting subpulses should be closed. It is natural to conclude that the circulating sparks would form
on average in a ring on the polar cap corresponding to a conal beam of radio emission originating at some height above
the polar cap \citep{gks93,kg98}.

The ``pulsar carousel model'' based on \EB\ drifting sparks is widely accepted by radio-observers but theorists usually
treat it with a great deal of reservation. Besides the binding energy problem (see \S4), they usually point out
problems with spark stability and memory, which is indeed necessary to explain the systematic subpulse drift within the
carousel model. This is a very important but difficult problem that requires sophisticated computer simulations. Here
we would only raise some phenomenological arguments. Following RS75, Gil \& Sendyk (2000; GS00 henceforth) argued that
both the characteristic spark dimension and the typical distance between the adjacent sparks are approximately equal to
the gap height. At any instant the polar cap should be populated by as many sparks as possible. Indeed, spark
discharges should develop wherever the accelerating potential drop exceeds the threshold for the magnetic pair
production. Such a maximum packing should stabilize an instantaneous arrangement of sparks on the polar cap surface.
Each spark develops and dies over a time scale of microseconds, so the question is what agency makes them to reappear
in the same place (such reappearance is necessary to explain the regular subpulse drifting). It seems that a natural
reason is thermal emission of charges at the base of each spark is heated by the return bombardment. Moreover, some
electrons and positrons from the outflowing spark plasma column would still be near the polar cap surface when the
returning potential drop again exceeds the pair formation threshold. These charged particles would initiate the very
next discharge \citep[see][and GS00 for more detailed discussion]{am98, mgp00}. Therefore, the only net motion of
sparks is the slow $\mathbf{E}\times\mathbf{B}$ drift across the planes of the surface magnetic field. For the
line-of-sight grazing the pulsar beam (corresponding to sparks operating at the polar cap boundary), this
circumferential drift motion will be observed as a systematic subpulse drift, while for a more central line-of-sight
trajectories only longitude-stationary modulation of the subpulse amplitude can be observed.

\section{Charge depleted inner acceleration region}

The inner acceleration region above the polar cap may be a result of deviation of the local charge density $\rho$ from
the co-rotational charge density \citep[][ GJ69 hereafter]{gj69}
\begin{equation}
\rho_{\rm GJ}=-\frac{{\mathbf\Omega}\cdot{\bf B}_{\rm s}}{2\pi
c}\approx\pm\frac{B_{\rm s}}{cP} ,
\label{roGJ}
\end{equation}
where the positive/negative sign corresponds to $^{56}_{26}$Fe ions/electrons. These two possibilities correspond to
antiparallel and parallel relative orientation of the magnetic and spin axes, respectively. As mentioned before, there
exists growing evidence \citep[see][ and references therein]{ug04} that the actual surface magnetic field $B_{\rm s}$
is highly non-dipolar. Its magnitude can be described in the form
\be B_{\rm s}=b B_{\rm d} \label{Bs}
\ee
\citep[][ GS00
hereafter]{gs00}, where the enhancement coefficient $b>1$ and
\be B_{\rm d}=2\times 10^{12}(P\dot{P}_{-15})^{1/2}~{\rm G}
\label{Bd}
\ee
is the canonical, star centered dipolar magnetic field, $P$ is the pulsar period in seconds and
$\dot{P}_{-15}=\dot{P}/10^{-15}$ is the period derivative. For the purpose of further considerations in this paper we
introduce here a convenient measure of the surface magnetic field \be \frac{B_{\rm s}}{B_{\rm
q}}=0.046(P\dot{P}_{-15})^{0.5}b \label{Bq} \ee where $B_{\rm q}=4.414\times 10^{13}$~G is the so called quantum
magnetic field \citep{e66}.

The polar cap is defined as the locus of the so-called open magnetic field lines that penetrate the light cylinder
(GJ69). Generally, the polar cap radius can be written as \be r_{\rm p}=1.45\times 10^4P^{-0.5}b^{-0.5}~{\rm cm}
\label{rp}
\ee (GS00), where the factor $b^{-0.5}$ describes squeezing of the polar cap area due to the magnetic flux conservation
in non-dipolar surface fields (as compared with the GJ69 dipolar configuration). One should realize, however, that the
polar cap radius $r_{\rm p}$ expressed by the above equation is only a characteristic dimension. In fact, in the
presence of strong non-dipolar surface magnetic field the actual polar cap can be quite irregular in shape. Also, the
light-cylinder radius and thus the canonical polar cap radius depends on the unknown particle inertia (e.g. Michel
1973). Therefore, $r_{\rm p}$ as expressed in equation (\ref{rp}) can be used as an order of magnitude estimate of the
actual polar cap radius.

The observationally deduced polar cap radii (see Section 3.2) are often much smaller than the canonical GJ69 value.
This seems consistent with strong non-dipolar surface magnetic field, that is large $b$ in equation (\ref{rp}). Given a
strong enough near-surface magnetic field (the required condition will be discussed below in \S\S4-7), charge depletion
in the acceleration region above the polar cap can result from binding of the positive $^{56}_{26}$Fe ions (at least
partially) in the neutron star surface. Positive charges then cannot be supplied at the rate that would compensate the
inertial outflow through the light cylinder. As a result, a significant part of the unipolar potential drop (GJ69,
RS75) develops above the polar cap, which can accelerate charged particles to relativistic energies and power the
pulsar radiation. The characteristic height $h$ of such an acceleration region is determined by the mean free path of
pair-producing high energy photons. In other words, the growth of the accelerating potential drop is limited by the
cascading production of an electron-positron plasma (e.g. RS75, CR80). The accelerated positrons would leave the
acceleration region, while the electrons would bombard the polar cap surface, causing a thermal ejection of ions, which
are otherwise more likely bound in the surface in the absence of additional heating. This thermal ejection would cause
partial screening of the acceleration potential drop $\Delta V$ corresponding to a shielding factor
\begin{equation}
\eta=1-\frac{\rho_{\rm i}}{\rho_{\rm GJ}},
\label{eta}
\end{equation}
where $\rho_{\rm i}$ is thermonically ejected charge density (see
also eq.[18]) and
\begin{equation}
\Delta V=\eta\frac{2\pi}{cP}B_{\rm s} h^2 ,
\label{deltafau}
\end{equation}
is the accelerating potential drop
\footnote{In this paper we ignore slight reduction of $\Delta V$ due to general
relativistic effect of frame dragging, which can affect the potential drop by as much as 27 percent (see Zhang et al.
2000; Gil \& Melikidze 2002).}, where $B_{\rm s}$ is defined by equation~(\ref{Bs}) and $h$ is the model dependent
height of the acceleration region (see section 5). The above expression for $\Delta V$ is the solution of Poisson
equation for a thin PSG, assuming that $\partial \eta/\partial h=0$ within the acceleration region.

The cascading sparking discharge in PSG works analogously to the scenario envisioned by RS75. The only difference is
that the accelerating potential drop is much smaller due to the partial screening by thermionic ions with charge
density $\rho_{\rm i}$ . Additional positive charge density $\rho_{+}$ is supplied by the spark produced positrons. The
potential drop is completely screened when the total charge density $\rho=\rho_{\rm i} + \rho_{+}$ reaches the
corotational value (eq.[\ref{roGJ}]). Then the spark plasma escapes from the region as discussed by \cite{am98}.
However, during a short escaping timescale $h/c\sim 10^{-8}$ s the thermionic ions still maintain the partial screening
as determined by equation (\ref{eta}). Thus, the returning potential drop quickly grows to the level defined by
equation (\ref{deltafau}) and the spark begins to develop again. This results in an intrinsically intermittent nature
of PSG and an unsteady particle flow into the magnetosphere, which has an important physical consequences for
generation of the coherent pulsar radio emission (see section 8).

In the stationary observer's frame the co-rotational charge density $\rho_{\rm GJ}=-{\bf\Omega}\cdot{\bf B}/2\pi c$ is
associated with the co-rotational electric field ${\bf E}_{\rm cor}({\bf r})=-({\bf\Omega}\times{\bf r})\times{\bf
B}({\bf r})/c$ and the magnetosphere co-rotates with the velocity ${\bf v}_{\rm cor}=c({\bf E}_{\rm cor}\times{\bf
B})/{B^2}$, where ${\bf B}$ is the magnetic field (dipolar) and ${\bf\Omega}$ is the pulsar spin axis
$(\Omega=2\pi/P)$. Now let us introduce a thin, charge depleted region extending to about $h$ above the polar cap
surface, with the accelerating potential drop described by equation (\ref{deltafau}). Since the charge density within
this region is lower than the co-rotating value $\rho_{\rm GJ}$, an additional electric field $\Delta{\bf E}$ appears
in this region, which makes the spark plasma to drift with a velocity ${\bf v}_{\rm d}={c}/{B^2}\left(\Delta
\mathbf{E}\times {\bf B}\right)$, as observed in the co-rotating frame. To estimate the values of $\Delta E$ let us
follow the original method of RS75 and consider the closed contour $abfea$ within the gap, as marked in Figure~4 of
RS75. Since the potential drop along segments $ab$ and $bf$ vanishes $(\Delta V_{ab}=\Delta V_{bf}=0)$, one then has
$\Delta V_{ae}=\Delta V_{fe}$, where $\Delta V_{ae}$ corresponds to the tangent electric field $\Delta{\bf E}$, while
$\Delta V_{ef}$ is just the acceleration potential drop expressed by equation (\ref{deltafau}). However, we can
consider another segment $f'e'$ (parallel to $fe$) and use the same argument to demonstrate that $\Delta V_{ae'}=\Delta
V_{ae}$ for any arbitrary pair of points $e$ and $e'$. This means that $\Delta E\approx 0$ within the RS75 gap, except
at the boundary region within about $h$ from the polar cap boundary, where $h$ is the height of the gap. The
perpendicular electric field $\Delta E$ grows rapidly from essentially zero to the corotational field ${\bf E}_{\rm
cor}$ in the boundary region. The situation can be slightly different if the gap height $h$ (length of segments $fe$
and $f'e'$) varies between the pole and the polar cap edge, although in the thin gap case of RS75 $(h\ll r_{\rm p})$
such variations cannot be large $(\Delta h/h\ll 1)$. In this case $\Delta V_{fe}\neq\Delta V_{f'e'}$ and $\Delta
V_{ae}\neq\Delta V_{ae'}$ and some residual $\Delta E$ can exist over the entire polar cap area. In any case, the
tangent electric field is strong only at the polar cap boundary where $\Delta E=0.5{\Delta
V}/{h}=\eta({\pi}/{cP})B_{\rm s}h$ (see Appendix~A in GMG03 for details). Here $B_{\rm s}$ is the surface magnetic
field (eqs.~[\ref{Bd}] and [\ref{Bs}]), $\Delta V$ is the accelerating potential drop (eq.~[\ref{deltafau}]), $h$ is
the gap height (eqs.~[\ref{TsTi}] and [\ref{Te}]), and $\eta$ is the shielding factor (eqs.[\ref{eta}], [\ref{eta2}]
and [\ref{Ti}]). This electric field causes that discharge plasma at the polar cap boundary performs a slow
circumferential motion with velocity $v_{\rm d}=c\Delta E/B_{\rm s}=\eta\pi h/P$. The time interval to make one full
revolution around the polar cap boundary is $\hat{P}_3\approx 2\pi r_{\rm p}/v_{\rm d}$. One then obtains the so called
tertiary drift periodicity
\begin{equation}
\frac{\hat{P}_3}{P}=\frac{r_{\rm p}}{2\eta h}, \label{P3P}
\end{equation}
which is the proper measure of the intrinsic drift rate, at least at the boundary of the polar cap. If the plasma above
the polar cap is fragmented into filaments (sparks) which determine the intensity structure of the instantaneous pulsar
radio beam, then, at least in principle, the tertiary periodicity $\hat{P}_3$ can be measured/estimated from the
pattern of the observed drifting subpulses \citep[e.g.][]{dr99,gs03}. In practice $\hat{P}_3$ is very difficult to
measure (mainly because of aliasing which is a severe problem even in pulsars with high signal-to-noise ratio, e.g.
DR99, GS03) and its value is known at the moment only in few cases. It is much easier to measure the primary drift
periodicity $P_3$, which in high signal-to-noise ratio is just a distance between the observed drift bands measured in
pulsar periods $P$ (there are also clever techniques that allow to measure $P_3$ even in cases with very low
signal-to-noise ratio; see Edwards \& Stappers 2002, WES06). Since $\hat{P}_3=NP_3$ (e.g. RS75), where $N\approx 2\pi
r_{\rm p}/2h$ is the number of sparks contributing to the drifting subpulse phenomenon observed in a given pulsar
(GS00), then one can write the shielding factor in the form
\begin{equation}
\eta\approx\frac{1}{2\pi}\frac{P}{P_3} ,
\label{eta2}
\end{equation}
which depends only on relatively easy-to-measure primary drift periodicity $P_3$ (WES06). This equation can be compared
with the definition of shielding factor (eq.[\ref{eta}]) in order to derive the amount of thermally ejected iron ions
or electrons (depending on the sign of charge of the polar cap). Apparently, the shielding parameter $\eta$ should be
much smaller than unity. Note that in the RS75 pure VG model one has $\eta=1$, which implies the predicted subpulse
drift at least an order of magnitude too fast as compared with observations. This means that the conditions within the
natural inner acceleration region above the polar cap should greatly differ from the pure vacuum gap proposed by RS75.

\subsection{Thermostatic self-regulation of the potential drop}
GMG03 argued that because of the exponential sensitivity of the accelerating potential drop $\Delta V$ to the surface
temperature $T_{\rm s}$, the actual potential drop should be thermostatically regulated (see also CR80). In fact, when
$\Delta V$ is large enough to ignite the cascading pair production, the back-flowing relativistic charges will deposit
their kinetic energy in the polar cap surface and heat it at a predictable rate. This heating will induce thermionic
emission from the surface, which will in turn decrease the potential drop that caused the thermionic emission in the
first place. As a result of these two oppositely directed tendencies, the quasi-equilibrium state should be
established, in which heating due to electron bombardment is balanced by cooling due to thermal radiation. This should
occur at a temperature $T_{\rm s}$ slightly lower than the critical temperature above which the polar cap surface
delivers thermionic flow at the corotational (GJ69) charge density level (see GMG03 for more details).

The quasi-equilibrium condition is $\sigma T_{\rm s}^4=\gamma m_{\rm e}c^3n$, where $\gamma=e\Delta V/m_{\rm e}c^2$ is
the Lorentz factor and $\Delta V$ is the accelerating potential drop (eq.~[\ref{deltafau}]). Here \be n=n_{\rm
GJ}-n_{\rm i}=\eta n_{\rm GJ}\label{n}\ee is the charge number density of back-flowing particles that actually heat the
polar cap surface, $\eta$ is the shielding factor (eq.~[\ref{eta}]), $n_{\rm i}$ is the charge number density of
thermally ejected flow and \be n_{\rm GJ}=\rho_{\rm GJ}/e=1.4\times 10^{11} b\dot{P}_{-15}^{0.5}P^{-0.5}~{\rm
cm}^{-3}\label{nGJ}\ee (eq.~[\ref{roGJ}]) is the corotational (GJ69) charge number density\footnote{Note that in the
original RS75 paper the charge number density corresponds to purely dipolar magnetic field ($b=1$), while in our
approach the surface magnetic field is highly non-dipolar ($b \gg 1$).}.  It is straightforward to obtain an expression
for the quasi-equilibrium polar cap surface temperature in the form
\be T_{\rm s}=\left( 2\times 10^6~{\rm K}\right) P^{-0.25} \dot{P}_{-15}^{0.25}
\eta^{0.5} b^{0.5} h_3^{0.5},
\label{Tsbis}
\ee
where $h_3=h/10^3$~cm is the normalized height of the acceleration region. This height is model dependent and we
discuss two possible models below (eqs.~[\ref{hCR}] or [\ref{hICS}]). Growing evidence suggests that in actual pulsars
$T_{\rm s}$ is few MK (see Table~1). Thus, from the above equation one can infer that $b\sim\eta^{-1}$, where the
shielding parameter should be much lower than unity (see discussion below eq.~[\ref{deltafau}]). As a consequence, the
enhancement coefficient $b=B_{\rm s}/B_{\rm d}$ (eq.~[\ref{Bs}]) should be much larger than unity (at least of the
order of 10). This is very consistent with the binding energy problem discussed in Section~4. Indeed, the extremely
high surface magnetic field helps to resolve this problem.

\subsection{Interrelationship between radio and X-ray signatures of drifting subpulses}
The predicted thermal X-ray emission luminosity from the polar cap with temperature $T_{\rm s}$ is $L_{\rm x}=\sigma
T_{\rm s}^4A_{\rm bol}$, where $A_{\rm bol}=\pi r_{\rm p}^2$ and $r_{\rm p}$ is the actual polar cap radius
(eq.~[\ref{rp}]) and $\sigma$ is the Stefan-Boltzmann constant. Thus $L_{\rm x}=1.2 \times
10^{32}(\dot{P}_{-15}/P^3)(\eta h/r_{\rm p})^2$~erg/s, which can be compared with the spin-down power
$\dot{E}=I\Omega\dot{\Omega}=3.95 I_{45}\times 10^{31}\dot{P}_{-15}/P^3$~erg/s, where $I=I_{45}10^{45}$g\ cm$^2$ is the
neutron star moment of inertia (in what follows we assume that $I_{45}=1$. We can now use the equation~(\ref{P3P}) in
the form $\eta h/r_{\rm p}\approx 0.5P/\hat{P}_3$ and derive the thermal X-ray luminosity from the polar cap heated by
sparks, i.e.
\be
L_{\rm x}=2.5\times
10^{31}\frac{\dot{P}_{-15}}{P^3}\left(\frac{\hat{P}_3}{P}\right)^{-2}
\label{Lx},
\ee
where $\hat{P}_3$ is the tertiary periodicity in drifting subpulses pattern (pulsar carousel time), which is equal to
the time interval needed for the discharge plasma to make one full revolution around the perimeter of the polar cap.

One can also derive the X-ray luminosity efficiency (with respect to the spin down luminosity)
\be \frac{L_{\rm x}}{\dot{E}}=0.63
\left(\frac{\hat{P}_3}{P}\right)^{-2} \label{LxE}.
\ee
We can see that equations (\ref{Lx}) and (\ref{LxE}) depend only on the observational data of the radio pulsars. It is
particularly interesting and important that both equations above do not depend on details of the sparking gap model
$(\eta, b, h)$. Thus, we have found a simple relationship between the properties of drifting subpulses observed in
radio band and the characteristics of X-ray thermal emission from the polar cap heated by sparks associated with these
subpulses. For PSR B0943$+$10, which is the only pulsar for which both $\hat{P}_3=37 P$, (DR99) and $L_{\rm x}=5 \times
10^{28}\ {\rm erg\ s}^{-1} \simeq 5\times 10^{-4} \dot E$, (ZSP05) are measured, the above equations hold very well. In
few other cases for which the circulational periodicity is measured $14<\hat{P}_3/P<37$ (see Table~1), which gives
$L_{\rm x}/\dot{E}\sim (1^{+4}_{-0.5})\times 10^{-3}$. Such correlation between the X-ray and the spin-down
luminosities is a well-known and intriguing property of rotation powered pulsars \citep{bt97,pcc02}. It has been
suggested that this correlation is a characteristic of a magnetospheric radiation \citep[e.g.][]{cgz98,zh00}. Here we
suggest that it can also be a characteristic property of the polar cap thermal radiation. Most likely, both mechanisms
contribute at a comparable level to the observed X-ray luminosity. However, one should realize that if the estimates
for PSRs 0809+74 and 0633+08 (Table 1) were correct, they would violate significantly the above correlation, so it
should be treated with caution until confirmed or discarded.

Using equations (\ref{P3P}) and (\ref{Tsbis}) we can write the polar cap temperature in the form
\be T_{\rm s}=(5.1\times
10^6{\rm K})b^{1/4}\dot{P}^{1/4}_{-15}P^{-1/2}\left(\frac{\hat{P}_3}{P}\right)^{-1/2} \label{Ts4},
\ee
where the enhancement coefficient $b=B_{\rm s}/B_{\rm d}\approx A_{\rm pc}/A_{\rm bol}$, $A_{\rm pc}=\pi r^2_{\rm pc}$
and $A_{\rm bol}=\pi r_{\rm p}^2$ (see Section~2). Since $A_{\rm bol}$ can be determined from the black-body fit to the
spectrum of the observed thermal X-ray emission, the above equations can also be regarded as independent of details of
the sparking gap model and depending only on the combined radio and X-ray data, similarly as equations (\ref{Lx}) and
(\ref{LxE}). This seems quite understandable, since it reflects the simple fact that within the sparking gap model both
the intrinsic drift rate and the polar cap heating rate are determined by the same electric field. Therefore, the
observational values of $L_{\rm x}$ and $\hat{P}_3/P$ should be related to each other through a combination of $P$ and
$\dot P$ that is proportional to the spin-down power $\dot E$, as it is really so in equations (\ref{Lx}) and
(\ref{LxE}). Let us summarize a set of assumptions that lead to these equations: (i) the electric field causing the
drift is estimated at the polar cap boundary\footnote{This is different from the original RS75 model, where the
electric field is estimated at the middle of the polar cap. Moreover, they used pure vacuum gap model.}
(eq.~[\ref{P3P}]), (ii) the gap is partially shielded (Eq.[\ref{n}]), and (iii) the surface magnetic field at the polar
cap is highly non-dipolar (eq.~[\ref{nGJ}]). If some or all of these assumptions were not valid, then the unknown
parameters such as $h$, $b$ and/or $\eta$ would appear in equations (\ref{Lx}) and (\ref{LxE}). Therefore, and
observational verification of these equations seems to be an easy way to confirm or discard the PSG model of the inner
accelerator in pulsars.

\subsection{Actual pulsars}
Table 1 presents the observational data and predicted values(computed from equations (\ref{Lx} - \ref{Ts4})) of a
number of quantities for five pulsars, which we believe show clear evidence of thermal X-ray emission from spark heated
polar caps. Besides the three cases with known $\hat{P}_3$ (B0943$+$10, B0826$-$34 and B0834$+$06), we also included
PSR B1133$+$16 (twin of PSR B0943$+$10) for which we estimated $\hat{P}_3$ and argued that its value was actually
measured but misinterpreted as the primary drift periodicity $P_3$ (shown in parenthesis in Table~1). PSR B0809+74, for
which an estimate of $\hat{P}_3$ exits, is included. We also added the Geminga pulsar (B0633+17) for which thermal
radiation from small polar cap was clearly detected.

{\it PSR B0943$+$10.} This is the best studied drifting subpulse radio pulsars with $P=1.09$~s, $\dot{P}_{-15}=3.52$,
$\dot{E}=10^{32}\ {\rm erg\ s}^{-1}$. DR99 clearly demonstrated that the observed subpulse drift is in this pulsar
aliased and found the alias-resolved values of drift periodicities $\hat{P}_3=37.4P$ and $P_3=1.86P$, which gave the
number of observed sparks $N=\hat{P_3}/P_3=20$. In an attempt to detect thermal X-ray signatures of these sparks ZSP05
observed this pulsar with the XMM-Newton. They obtained a spectrum consistent with thermal BB fit with a bolometric
luminosity $L_{\rm x}\approx 5\times 10^{28}\ {\rm erg\ s}^{-1}$ and a polar cap surface area $A_{\rm bol}=10^7(T_{\rm
s}/3\times 10^6{\rm K})^{-4}{\rm cm}^2\sim(1^{+4}_{-0.4})\times 10^7\ {\rm cm}^2$, which was much smaller than the
conventional polar cap area $A_{\rm pc}=6\times 10^8\ {\rm cm}^2$. This corresponds to the best fit temperature $T_{\rm
s}=3.1\times 10^6$~K (see Fig.~1 in ZSP05). The predicted values of $L_{\rm x}$ and $L_{\rm x}/\dot{E}$ calculated from
equation~(\ref{Lx}) and (\ref{LxE}), respectively, agree very well with the observational data. The surface temperature
$T_{\rm s}$ calculated from equation~(\ref{Ts4}) with $b=A_{\rm pc}/A_{\rm bol}$ is also in very good agreement with
the best fit. The shielding parameter $\eta\approx 0.1$ is derived from equation~(\ref{eta2}).

{\it PSR B1133$+$16.} This pulsar with $P=1.19$~s, $\dot{P}_{-15}=3.7$, and $\dot{E}=9\times 10^{31}\ {\rm erg\
s}^{-1}$ is almost a twin of PSR B0943$+$10. KPG05 observed this pulsar with Chandra and obtained a spectrum consistent
with thermal BB fit $L_{\rm x}=6.7\times 10^{28}\ {\rm erg\ s}^{-1}$ (we used their $L_{\rm
x}/\dot{E}=3.6^{+0.6}_{-0.7}\times 10^{-4}<\cos\theta>^{-1}$ with $<\cos\theta>=0.47$), $A_{\rm
bol}=(0.5^{+0.5}_{-0.3})\times 10^7~{\rm cm}^2$ and $T_{\rm s}\approx 2.8\times 10^6$~K. These values are also very
close to those of PSR B0943$+$10, as should be expected for twins. Using equations (\ref{Lx}) or (\ref{LxE}) we can
predict $\hat{P}_3/P=27^{+5}_{-2}$ for B1133+16. Interestingly, \citet[][ R86 henceforth]{r86} gives the primary drift
periodicity $P_3=(5.3\pm 1.2)P$ for the conal components (probably aliased), but in the saddle between these components
R86 reports $P_3=(25\pm 3)P$. However, recently WES06, reported in this pulsar $P_3/P=3\pm 2$, as well as long period
drift feature with $(33\pm 3)P$ in the trailing profile component. We thus claim that this long period (shown in
parenthesis in Table 1) is the actual tertiary periodicity $\hat{P}_3/P=33^{+3}_{-3}$, which agrees very well with our
prediction $\hat{P}_3/P=27^{+5}_{-2}$. This interpretation is strongly supported by Fig.~10 in \citet{n96}, where the
fluctuation spectrum of PSR B1133+16 shows prominent long period feature at 0.031 cycles/period, corresponding to 32
pulsar periods. It is worth noting that $\hat{P}_3/P_3=33\pm 3$ is very close to 37.4 measured in the radio twin PSR
B0943+10. Note also that the number of sparks predicted from our hypothesis is $N=\hat{P}_3/P_3=(33\pm 3)/(3\pm
2)=11^{+25}_{-6}$, thus it is quite possible that actually $N$ is close to 20, as in the case of twin pulsar
B0943$+$10.

{\it PSR B0826$-$34.} With $P=1.85$~s, $\dot{P}_{-15}=1.0$, $\dot{E}=6\times 10^{30}\ {\rm erg\ s}^{-1}$, this pulsar
has $P_3\approx 1P$ (highly aliased) and $\hat{P}_3\approx 14P$, i.e. $N=14$ \citep{ggk04,elg05}. With these values we
can predict from equation~(\ref{Lx}) that $L_{\rm x}=2\times 10^{28}\ {\rm erg\ s}^{-1}$. This pulsar should be as
bright as PSR B0943$+$10 and thus it is worth observing with {\em XMM-Newton}. The shielding parameter $\eta\approx
0.16$ (eq.~[\ref{eta2}]).

{\it PSR B0834$+$06.} With $P=1.27$~s, $\dot{P}_{-15}=6.8$, and $\dot{E}=1.3\times 10^{32}\ {\rm erg\ s}^{-1}$, this
pulsar has $\hat{P}_3=15P$, and $P_3=2.16\ P$ \citep{ad05}, implying the number of sparks $N\approx 15/2.16\approx 7$.
From equation~(\ref{Lx}) we obtain $L_{\rm x}=37\times 10^{28}\ {\rm erg\ s}^{-1}$. This pulsar should be almost 8
times more luminous than PSR B0943$+$10, and we strongly recommend to observe it with {\em XMM-Newton}. The shielding
parameter $\eta\approx 0.07$ (eq.~[\ref{eta2}]).

{\it PSR B0809+74.} This is one of the most famous pulsars with drifting subpulses, with $P=1.29$~s,
$\dot{P}_{-15}=0.17$, $\dot{E}=3.1\times 10^{30}\ {\rm erg\ s}^{-1}$ and well determined value of the primary drift
period $P_3=11\ P$. Recently, \citet{lsr03} estimated the circulational tertiary period $\hat{P}_3 \ge 150P$. This is
very long as compared with other pulsars considered above, but see the Geminga case below. Taking the lower limit
$\hat{P}_3=150\ P$ we can predict from eq.~(\ref{Lx}) thermal X-ray luminosity at the level $L_{\rm x}=3\times 10^{25}\
{\rm erg\ s}^{-1}$, which is a very low compared with other nearby pulsars. The efficiency ratio $L_{\rm x}/\dot{E}=3
\times 10^{-6}$. The shielding factor $\eta =0.014$ (eq.~[\ref{eta2}]).

{\it PSR B0633+17 (Geminga).} This pulsar with $P=0.237$ s and $\dot{P}_{-15}=10.97$ is a problematic radio emitter,
but well known for pulsating optical, X-ray and $\Gamma$-ray emission. Recently, \citet{cetal04} detected thermal
emission from an ~60 meter-radius polar cap heated to $T_{\rm s}=(1.9 \pm 0.3) \times 10^6$ K, with the bolometric
thermal luminosity $L_{\rm x}=1.5 \times 10^{29}\ {\rm erg\ s}^{-1}$. Since the canonical radius of the polar cap is
300 m, then the amplification factor $b(300/60)^2=25$ (eqs.~[\ref{Bs}] and [\ref{rp}]). Using the observed value of
$L_{\rm x}$ we obtain from equation (13) the predicted value of $\hat P_3/P=400$, which is even longer than the minimum
estimate in PSR B0809+74. Now taking this value and the estimate for $b$ we obtain from equation (\ref{Ts4}) $T_{\rm
s}=2.13 \times 10^6$ K, which is in very good agreement with observational range of the surface temperature $(1.9 \pm
0.3) \times 10^6$ K. The efficiency ratio $L_{\rm x}/\dot{E}=4 \times 10^{-6}$, close to the estimate in PSR B0809+74.

\citet{bwt04} revealed the X-ray emission from a number of old rotation-powered pulsars. Among them is PSR B0823+26,
which is well known to exhibit regularly drifting subpulses (see Table 1) with $P_3=(2.2 \pm 0.2)P$ (WES06). Such an
old drifting pulsar is an ideal case to study the spark associated thermal radiation from hot polar cap. \citet{bwt04}
found an upper limit of the thermal contribution from a hot polar cap as being $T_{\rm pc}=1.17 \times 10^6$~K, which
seems low as compared with estimates given in Table 1. However, \citet{bwt04} used a full canonical polar cap with
radius 299 meters to estimate $T_{\rm pc}$. In the actual non-dipolar surface magnetic field one should re-scale
$T_{\rm s}=b^{0.25} T_{\rm pc}$, where $b=(r_{\rm p}/r_{\rm pc})^2=B_{\rm s}/B_{\rm d}$ (eqs.~[\ref{Bs},\ref{rp}]).
Since $B_{\rm d}=10^{12}$~G in this case, $b$ would be at least 10. This raises the lower limit of $T_{\rm pc}$ to be
above $2 \times 10^6$~ K, consistent with our model. The same applies to another old pulsar B0950+08 studied by
\citet{bwt04}.

\section{Binding energy problem}

The phenomenon of drifting subpulses seems to indicate strongly a presence of the charge depleted acceleration region
just above the polar cap. However, the existence of such region strongly depends on the binding energy of
$^{56}_{26}$Fe ions, which are the main building blocks of neutron star surface \citep{um95,l01}. If this cohesive
energy is large enough to prevent thermionic emission, a charge depleted acceleration region can be formed above the
polar cap. Normally, at the solid-vacuum interface, the charge density of outflowing ions is roughly comparable with
density of the solid at the surface temperature $kT_{\rm s}=\varepsilon_c$, where $\varepsilon_c$ is the cohesive
(binding) energy and $k=8.6\times 10^{-8}$~keVK$^{-1}$ is the Boltzman constant. However, in the case of pulsars, only
the corotational charge density $\rho_{\rm GJ}$ can be reached, and the $^{56}_{24}$Fe ion number density corresponding
to $\rho_{\rm GJ}$ is about $\exp(-30)$ times lower than in the neutron star crust. Since the density of outflowing
ions $\rho_{\rm i}$ decreases in proportion to $\exp(-\varepsilon_c/kT_{\rm s})$, one can then write $\rho_{\rm
i}/\rho_{\rm GJ}\approx\exp(30-\varepsilon/kT_{\rm s})$. At the critical temperature \be T_{\rm
i}=\frac{\varepsilon_c}{30k} \label{Ti}
\ee the ion outflow reaches the maximum value $\rho_{\rm i}=\rho_{\rm GJ}$ (Eq.~[\ref{roGJ}]), and the numerical coefficient
30 is determined from the tail of the exponential function with an accuracy of about 10\%.

Calculations of binding energies are difficult and uncertain (see Usov \& Melrose 1995, 1996 and Lai 2001 for detailed
discussion). In this paper we use the results of \citep[ J86 henceforth]{j86}, which were recommended by \citet{l01} in
his review paper as more robust than others. J86 obtained $\varepsilon_c$=0.29, 0.60 and 0.92 keV for $B_{\rm s}=2,5$
and $10\times 10^{12}$~G, respectively. These values can be approximately represented by the function
$\varepsilon_c\simeq(0.18{\rm keV})(B_{\rm s}/10^{12})^{0.7}$~G. Using equations (\ref{Bs}), (\ref{Bd}) and (\ref{Ti}),
this could be converted into the critical temperature
\be
T_{\rm i}\simeq \left( 0.7\times 10^5~{\rm K} \right)\left(\frac {B_{\rm s}}{10^{12}~{\rm G}}\right)^{0.7}\simeq \left(
1.2\times 10^{5}~{\rm K}\right) b^{0.7} \left( P\dot{P}_{-15}\right)^{0.36} \approx \left(10^6~{\rm K}\right)
\left(\frac{ B_{\rm s}}{B_{\rm q}}\right)^{0.7} , \label{Tibis}
\ee
where we use the ratio $B_{\rm s}/B_{\rm q}$ (eq.~[\ref{Bq}]) for the convenience of presenting the results in the
graphical form (Fig.~1 and 2). Above this temperature the thermionic ion flow reaches the maximum GJ density at the
surface, and the polar cap flow will be space charge limited. An acceleration potential can be still developed but the
growth rate is slow compared with the PSG model we are developing in this paper. At temperatures below $T_{\rm i}$
charge-depletion would happen right above the surface, and an efficient acceleration region would form, which should be
discharged in a quasi-steady manner by a number of sparks, similar to the original suggestion of RS75 for the pure
vacuum gap case (see more detailed discussion in GMG03). The electron-positron plasma produced by sparking discharges
co-exists with the thermally ejected ions, whose charge density can be characterized by the shielding factor (defined
by eq.~[\ref{eta}]) in the form
\be
\eta=1-\exp\left[ 30\left(1-\frac{T_{\rm i}}{T_{\rm s}}\right)\right]. \label{etabis}
\ee
As one can see from equations~(\ref{eta}) and (\ref{etabis}), at the temperature $T_{\rm i}=T_{\rm s}$ the shielding
factor $\eta=0$ (corresponding to fully developed space-charge limited flow with $\rho_{\rm i}=\rho_{\rm GJ}$), but
even a very small drop of $T_{\rm s}$ below $T_{\rm i}$, much smaller than 10\%, corresponds to creation of the pure
vacuum gap with $\eta=1$ ($\rho_{\rm i}=0$). Thus, the condition for partially screened charge depleted acceleration
region can be written in the form $T_{\rm s}\lesssim T_{\rm i}$, meaning that the actual surface temperature $T_{\rm
s}$ should be slightly lower (few percent) than the critical ion temperature $T_{\rm i}$, which for a given pulsar is
determined purely by the surface magnetic field $B_{\rm s}$ (eq.~[\ref{Tibis}]). Practically, one can use
\be
T_{\rm s}=T_{\rm i},
\label{TsTi}
\ee
to denote the condition of forming a PSG, with the caveat that in reality $T_{\rm s}$ cannot be exactly equal to
$T_{\rm i}$ but should be a few percent lower.

As mentioned above, the polar cap surface can be, in principle, negatively charged $({\mathbf\Omega}\cdot{\bf B}>0)$.
In such a case (called ``antipulsars'' by RS75) the polar cap surface can deliver an electron flow. Following GMG03 we
assume that this electron flow is also determined mainly by thermoemission, with the corresponding shielding factor
$\eta=1-\rho_{\rm e}/\rho_{\rm GJ}=1-\exp[25(1-T_{\rm e}/T_{\rm s})]$, where $\rho_{\rm e}$ is the charge density of
thermionic electrons. The critical electron temperature is
\be
T_{\rm e}\simeq(5.9\times 10^5~{\rm
K})b^{0.4}P^{0.16}\dot{P}^{0.2}_{-15}\sim(10^6~{\rm K})\left(
\frac{B_{\rm s}}{B_{\rm q}}\right)^{0.4}
\label{Te}
\ee
(see GMG03 for more details), and in analogy to equation~(\ref{TsTi}) the condition for creation of charge depleted
acceleration region is $T_{\rm s}=T_{\rm e}$. Since the enhancement coefficient $b>>1$ (see the discussion following
eq.~[\ref{Tsbis}]) so that $B_{\rm s}/B_{\rm q}\sim 1$ (eq.~[\ref{Bq}]), both $T_{\rm i}$ and $T_{\rm e}$ can be close
to $T_{\rm s}\sim$ a few MK. Since $T_{\rm i}$ and $T_{\rm e}$ are similar, we will only include the ion case
(eq.~[\ref{Tibis}]) in the following discussion, keeping in mind that our considerations are general and independent of
the sign of the polar cap charge, at least qualitatively.

 \section{PSG models of the inner acceleration region in pulsars}

So far, we have considered the macroscopic properties of the inner accelerator region in a way independent of details
of the sparking gap model. Now we begin to discuss the microscopic properties of this model. In particular, we will
present results of model calculations similar to those presented by GM02 but for a much larger sample of drifting
subpulse pulsars and with the inclusion of partial screening due to thermionic emission from the polar cap surface. The
condition for the formation of the gap used in GM02 was $T_{\rm c}/T_{\rm s}>1$, while in this paper we used $T_{\rm
c}/T_{\rm s}=1$ (eq.~[\ref{TsTi}]) as a result of the thermostatic regulation considered by GMG03.

Growth of the accelerating potential drop (eq.~[\ref{deltafau}]) is limited by the cascading production of relativistic
electron-positron pair plasma in strong and curved magnetic field. RS75 derived a famous formula for their VG height
assuming that the quasi-steady breakdown is driven by magnetic pair production induced by the curvature radiation seed
photons. They used \citet{e66} approximation, which is valid in relatively low magnetic fields $B_{\rm s}/B_{\rm
q}\lesssim 0.1$, which is not relevant for the magnetic field at the polar cap considered in this paper. In the strong
surface magnetic field, i.e. $B_{\rm s}>5\cdot 10^{12}$~G, the high energy photons with energy $E_f=\hbar\omega$
produce electron-positron pairs at or near the kinematic threshold $\hbar\omega=2mc^2/\sin\theta$, where
$\sin\theta=l_{\rm ph}/{\cal R}$, $l_{\rm ph}$ is the photon mean free path for pair formation and ${\cal R}={\cal
R}_6\cdot 10^6$~cm is the radius of curvature of magnetic field lines. This regime is called the near threshold (NT)
conditions \citep[e.g.][]{dh83}. Two VG models can be considered under the NT conditions: Curvature Radiation dominated
Near Threshold Partially Screened Gap (CR-NTPSG) model and Inverse Compton Scattering dominated Near Threshold
Partially Screened Gap (ICS-NTPSG) model, in which the potential drop is limited by pair production of the CR and the
resonant ICS seed photons, respectively. Proper expressions corresponding to pure vacuum gap case have been derived by
GM01 and GM02. Below we give generalized formulae by including partial screening due to the thermal flow from the polar
cap surface (GMG03 and references therein).

\vspace{0.5cm} \noindent{\it CR-NTPSG model}

In this model the cascading $e^-e^+$ pair plasma production is driven (or at least dominated) by the curvature
radiation photons with typical energy $\hbar\omega=(3/2)\hbar\gamma^3c/{\cal R}$. where $\gamma$ is the typical Lorentz
factor of electrons/positrons moving relativistically along the local surface magnetic field lines with a radius of
curvature ${\cal R}$. In the quasi-steady conditions the height $h$ of the acceleration region is determined by the
mean free path that is $h\sim l_{\rm ph}$ for pair production by energetic CR photon in the strong and curved magnetic
field. Following GM01 (see also GM02) and including the partial screening effect (GMG03) we obtain
\be h\approx h^{\rm CR}=(3.1\times 10^{3}){\cal
R}_6^{2/7}\eta^{-3/7}b^{-3/7}P^{3/14}\dot{P}_{-15}^{-3/14}~{\rm
cm}. \label{hCR} \ee

\vspace{0.5cm} \noindent{\it ICS-NTPSG model}

In this model the cascading pair plasma production is driven (or at least dominated) by the resonant inverse Compton
scattering \citep[e.g.][]{zqlh97}, with typical photon energy $\hbar\omega=2\gamma\hbar eB_{\rm s}/mc$. In the
quasi-steady conditions the height $h$ of the acceleration region is determined by the condition $h \sim l_{\rm e}$
where $l_{\rm e}=0.0027\gamma^2(B_{\rm s}/10^{12}{\rm G})^{-1}(T_{\rm s}/10^6~{\rm K})^{-1}$ is the mean free path of
the electron to emit this high energy ICS photon (ZHM00), $B_{\rm s}$ is the surface magnetic field and $T_{\rm s}$ is
the actual surface temperature. Following GM01 and including partial screening (GMG03) we obtain
\begin{equation}
h\approx h^{\rm ICS}=(5\times 10^3){\cal
R}_6^{4/7}\eta^{-1/7}b^{-1}P^{-5/14}\dot{P}_{-15}^{-1/2}~{\rm cm}.
\label{hICS}
\end{equation}
In the above equations ${\cal R}_6={\cal R}/10^6$~cm is the normalized curvature radius of the surface magnetic field
lines.

Besides resonant ICS, also the non-resonant ICS with a characteristic energy $\hbar\omega \sim \max(\gamma^2 k T,
\gamma m c^2)$ has been widely discussed in literature. This process usually dominates the gap dynamics in pulsars with
moderate and low magnetic fields \citep{zqlh97,ha01,hm02}. Since the polar caps in our model typically anchor a very
strong curved magnetic field, the non-resonant ICS is unimportant and we ignore it in our detailed treatment.
Throughout this paper ICS refers to resonant ICS, and the equations (\ref{hCR}) and (\ref{hICS}) will be referred as to
CR and ICS cases, respectively.

 \subsection{CR-NTPSG}

The results of model calculations for 102 pulsars with drifting subpulses (see Table~1) are presented in Figure~1 for
the ion $T_{\rm i}$ critical temperature (eq.~[\ref{Tibis}]). We have plotted $B_{\rm s}/B_{\rm q}$ (left-hand side
vertical axes) versus the pulsar number (which corresponds to a particular pulsar according to Table 1). The actual
value of $B_{\rm s}/B_{\rm q}$ (eq.~[\ref{Bq}]) for a given pulsar was computed from the condition $T_{\rm s}=T_{\rm
i}$, where $T_{\rm s}$ is the actual surface temperature (eq.~[\ref{Tsbis}]), with the height $h$ of CR driven
acceleration region determined by equation~(\ref{hCR}). This condition leads to the expression $B_{\rm s}/B_{\rm
q}=126.4P^{-20/29}{\cal R}_6^{10/29}\eta^{20/29}$, which allows to sort pulsars according to decreasing value of pulsar
period $P$, which is marked on the top of the Figure~1. The vertical axes on the right hand side are expressed in terms
of the surface temperature $T_{\rm s}$ computed from equation~(\ref{Tibis}). Different panels correspond to different
normalized radii of curvature ranging from ${\cal R}_6=0.1$ to ${\cal R}_6=0.009$. Different symbols used to plot
exemplary curves correspond to arbitrarily chosen values of shielding factor ranging from $\eta=0.015$ to $\eta=0.15$
(each curve represents the same shielding conditions $\eta$ for all 102 pulsars considered).

The vertical line in Figure~1 corresponds to PSR B0943$+$10 (N1=41 in Table~1), which was observed using XMM-Newton by
ZSP05. Three horizontal lines correspond to $T_{\rm s}$ equal to about 2, 3 and 4~MK (from bottom to the top),
respectively, calculated from equation~(\ref{Tibis}). (More exactly, we marked $T_{\rm s}=T_{\rm i}=2.08$, 3.11 and
4.14 for $B_{\rm s}/B_{\rm q}=2.7$, 4.8 and 7.2, respectively). Thus, the hatched areas encompassing these lines
correspond to the range of surface temperatures $T_{\rm s}\cong(3\pm 1)\times 10^6$~K deduced observationally for the
polar cap of B0943+10 from XMM-Newton observations (ZSP05, see their Fig.~1).

\subsection{ICS-NTPSG model}

Let us now consider the resonant inverse Compton scattering radiation model (ICS-NTPSG). The major difference between
CR and ICS cases is the additional regulation of $T_{\rm s}$ caused by the condition $h \approx l_{e}$, where
$l_{\rm{e}}$ is inversely proportional to $T_{\rm s}$. In fact, the increase of the surface temperature causes the
decrease of $h$, and hence, the decrease of $\Delta V$. This makes the back-flow heating of the polar cap surface less
intense. Let us estimate the surface temperature in the ICS dominated case. The average Lorentz factor of electrons or
positrons can be estimated by the gap height and the partially screened potential drop as $\gamma =\left(1.1\times
10^{2}\right) P^{1/6} \dot P_{-15}^{-1/6} b^{1/3}(T_{\rm s}/10^6)^{2/3}\eta^{-1/3}$. Using the kinematic near threshold
condition and the expression of the resonant ICS photons energy (see section~2) we get $\hbar\omega=\left( 7.5\times
10^{-8}\right) \gamma b ( P \dot P_{-15})^{-1/2}$, which leads to another estimate for the average Lorentz factor
$\gamma =2.5 \left(T_{\rm s}/10^6\right)^{1/3} {\cal R}_{6}^{1/3}$. Combining these two expressions for $\gamma$ we
find that for the case of the ICS dominated NTPSG the following relationship should hold $\eta=8.14\times
10^{-5}P^{1/2}\dot P_{-15}^{-1/2}b \left(T_{\rm s}/10^6\right){\cal R}_{6}^{-1}$. As a result, contrary to the CR case,
the shielding factor $\eta $ is not a free parameter in the ICS case, and the analysis similar to that presented in
Figure~1 is not relevant. However, we can use the quasi-equilibrium condition $\sigma T_{\rm s}^{4}=\gamma
m_{e}c^{3}n_{0}\eta$, which leads to $T_{\rm s}=\left( 8.8\times 10^{4}\right) \left( \gamma \eta b P^{-3/2}\dot
P_{-15}^{-1/2}\right)^{1/4}$. Using the expressions of $\eta$ and $\gamma$ we can then obtain
\be
T_{\rm s}=(1.5\times 10^4~{\rm K}){\cal R}_{6}^{-1/4} P^{-3/8}\dot P_{-15}^{-3/8} b^{3/4}. \label{ics3}
\ee
Consequently, we can obtain the expression for the shielding factor
\be
\eta=\left(
1.2\times 10^{-6}\right) P^{1/8} \dot P_{-15}^{-7/8}{\cal
R}_{6}^{-5/4} b^{7/4}.
\label{ics4}
\ee
Thus, for a given pulsar $(P, \dot{P})$ the values of $T_{\rm s}$ and $\eta$ are determined by the parameters of the
local surface magnetic field ${\cal R}_6$ and $b$. In Figure~2 we plot $T_{\rm s}$ as a function of $B_{\rm s}/B_{\rm
q}$ (eq.~[\ref{Bq}]) for different values of ${\cal R}_6$ and $\eta$. The smooth lines in both panels correspond to the
ion critical temperature, while other lines marked by different symbols shown in legends correspond to the actual
surface temperature calculated for the CR case (upper panel) and the ICS case (lower panel). We look for partially
screened solutions corresponding to the intersection of smooth and symbol-broken lines.

As seen from the lower panel of Figure~2, equations (\ref{ics3}) and (\ref{ics4}) cannot be self-consistently
satisfied. In fact, unlike in the CR case presented in the upper panel, in the ICS case the surface temperature $T_{\rm
s} \simeq 10^6$~K (eq.~\ref{ics3}) is considerably lower than $T_{\rm i}$ (eq.~\ref{Tibis}). Bringing $T_{\rm s}$ close
to $T_{\rm i}$ would imply $\eta=1$ (pure vacuum), which is inconsistent with equation \ref{ics4}. Thus, if the pair
creation is controlled by ICS, then the acceleration region should be a pure vacuum gap rather than a PSG. This is
consistent with previous results of GM02, who concluded that in the pure vacuum gap case ICS seed photons are more
efficient in driving the cascading pair production. However, GM02 failed to understand that CR pair production
mechanism produces more polar cap heating and thus makes the gap partially screened. In such a case the ICS pair
production still occurs but it is no longer the dominant mechanism to control the properties (height, potential drop,
etc) of the acceleration region above the polar cap.

\section{Thermal signatures of hot polar cap}

Knowing that ICS cannot contribute significantly to the pair production process in a PSG, let us consider polar cap
heating by back-flow plasma particles created from CR seed photons. Using the near threshold condition in the form
$h=(4/3)(mc{\cal R}^2/\hbar\gamma^3)=h_310^3$~cm (see below eq.~[\ref{Bq}]) we can rewrite equation~(\ref{Tsbis}) in
the form
\be
T_{\rm s}=(7.7\times 10^6{\rm K})P^{-2/7}\eta^{2/7}{\cal R}_6^{1/7}\left(\frac{B_{\rm s}}{B_{\rm q}}\right)^{2/7} .
\label{Ts}
\ee
Consequently, we can obtain efficiency of thermal X-ray emission in the form
\be
\frac{L_{\rm x}}{\dot{E}}=0.15P^{19/14}\dot{P}_{-15}^{-1/2}\eta^{8/7}{\cal R}_6^{4/7}\left(\frac{B_{\rm s}}{B_{\rm
q}}\right)^{1/7} , \label{LxE2}
\ee
where $L_{\rm x}=\sigma T_{\rm s}A_{\rm bol}$ is the X-ray luminosity from the hot polar cap with the surface area
$A_{\rm bol}=\pi r_{\rm p}^2$ (eq.~[\ref{rp}]) and $\dot{E}=I\Omega\dot{\Omega}=3.95\times
10^{31}\dot{P}_{-15}/P^3$~erg/s is the spin-down power. For the twin radio pulsars PSR B0943$+$10 and PSR B1133$+$16
the X-ray signatures of the hot polar cap are $T_{\rm s}\sim 3\times 10^6$~K, $\dot{E}\approx 10^{32}$~erg/s and
$L_{\rm x}/\dot{E}\sim 5\times 10^{-4}$ (see Table~1). Using these measurements we can infer from equations (\ref{Ts})
and (\ref{LxE2}) an approximate condition in the form $\eta^2{\cal R}_6\sim 6.5\times 10^{-5}$. Since in PSR B0943+10
$P_3/P=1.86$ then according to equation (9) we obtain the normalized radius of curvature ${\cal R}_6=0.009$ in this
pulsar (see next section). The microscopic equations (26) and (26) along with two macroscopic equations (14) and (15)
provide full knowledge about the spark driven inner acceleration regions in pulsars. The PSG model behind these
equations should be easily verifiable because they are either completely independent or weakly dependent on physical
parameters of the acceleration region.

Figures~3 and 4 present calculations of the efficiency $L_{\rm x}/\dot{E}$ and the temperature $T_{\rm s}$,
respectively, for 102 pulsars from Table~1. One should emphasize that these calculations utilized the two above
equations and did not assume a priori an existence of the PSG in form of the condition $T_{\rm s}=T_{\rm c}$
(eqs.~[\ref{Tibis}] and [\ref{TsTi}]). Thus, these figures should be considered in association with Figure~1, which
does take into account this condition. The vertical solid lines correspond to PSR B0943$+$10 for which $T_{\rm s}=(3\pm
1)\times 10^6$~K (corresponding to $B_{\rm s}/B_{\rm q}$=4.8 in Fig. 1) and $L_{\rm x}/\dot{E}=(5\pm 2)\times 10^{-4}$
(ZSP05). Due to weak dependence of the efficiency on the magnetic field we decided to use only one value of $B_{\rm
s}/B_{\rm q}$=4.8 in Figure~3 to avoid overlapping of too many lines. The observational range of $L_{\rm x}/\dot{E}$
and $T_{\rm s}$ are marked by the hatched belts in Figsure~3 and 4, respectively. Since $L_{\rm x}/\dot{E}$ in
equation~(\ref{LxE2}) depends on both $P$ and $\dot{P}$ we have sorted pulsars in Figure~3 according to increasing
value of $\dot{E}$, which are marked on the top of the figure (N2 in Table~1). If $L_{\rm x}/\dot{E}\sim 10^{-3}\div
10^{-4}$ for all pulsars, then one can say that the shielding factor $\eta$ should be larger for pulsars with larger
$\dot{E}$. Also, it follows from Figure~4 that if $T_{\rm s}=(2\div 4)\times 10^6$~K in all pulsars, then the shielding
factor $\eta$ should be larger for longer period pulsars. This is consistent with equation~(9).

\section{Special case of PSR B0943$+$10}

The microscopic parameters of PSR B0943$+$10 are described in section 3.2 (with appropriate entry in Table 1). Let us
now discuss this special case in more detail. PSR B0943$+$10 is the only pulsar for which both $\hat{P}_3$ and $L_{\rm
x}$ are known observationally. As already mentioned above, both the subpulse drift rate and the polar cap heating rate
(due to subpulse-producing sparks) are related to the same value of the accelerating potential drop. In section 3, we
found a ``clear cut'' formula involving two observables: the X-ray luminosity $L_{\rm x}$ and the tertiary drift
periodicity $\hat{P}_3$, which is independent of microscopic details of the acceleration region. This relationship
holds very well for PSR B0943$+$10 (see Table 1). It is important to check whether this "model independent" approach is
consistent with detailed model calculations, involving microscopic conditions prevailing in PSG. According to equations
(\ref{P3P}) and (\ref{eta2}) we have the shielding parameter $\eta=0.09$ and the complexity parameter $r_{\rm
p}/h=N/\pi=6.36$, in consistency with GS00 \citep[see also][]{gmm02}. Let us analyze in this respect our Figure~1,
which allows to read off the physical parameters of the acceleration region for a given pulsar within a particular
model $\eta$. For PSR B0943$+$10, only the intersections of the vertical lines corresponding to N1=41 (see Table~2)
with the hatched belts, or even with the horizontal line $T_{\rm s}=$ 3MK, are relevant. The above value of $\eta=0.09$
corresponds to the radius of curvature ${\cal R}_6\sim 6.5\times 10^{-5}\eta^{-2}\sim 0.009$ (see the condition below
eq.~[\ref{LxE2}]), and thus from the lower-right panel of Figure~1 we have $B_{\rm s}/B_{\rm q}\sim 5$ or $B_{\rm
s}\sim 2\times 10^{14}$~G and $T_{\rm s}\sim 3\times 10^6$~K. This set of parameters can be also inferred from the
upper panel in Figure~2. These values therefore are consistent with the appropriate entry for PSR B0943+10 in Table 1.
This consistency between macro- and micro-scopic calculations holds also for other pulsars from Table 1, as illustrated
in Figure~3. This set of parameters can be also inferred from the upper panel in Figure~2.

The surface temperature about 3MK (ZSP05), which implies $B_{\rm s}=bB_{\rm d}\sim 2\times 10^{14}$~G if one applies
J86 theory (eq.~[\ref{Tibis}]) to the case of B0943+10. The value exceeding $10^{14}$~G may seem extremely high, but at
least three radio pulsars have dipolar magnetic fields above $10^{14}$~G \citep[][ and references therein]{Mc03}. The
non-dipolar surface magnetic fields could be even stronger. The reason that the inferred non-dipolar field is even
stronger than in the pure vacuum case found in GM02 for the ICS seed photons is that we are using a lower binding
energy calculations of \citet{j86}, which are more robust than those of \citet{as91} \citep[see][ for critical
review]{l01}. However, generally the surface magnetic field required to form CR-NTVG obtained by GM02 is still higher
than that for CR-NTPSG obtained in this paper.

In section 3.1 we assumed a pure black-body conditions at the polar cap. GM02 considered deviations from the black-body
conditions by introducing the so-called heat-flow coefficient $\kappa=Q_{rad}/(Q_{rad}+Q_{cond})$, which described the
amount of heat conducted beneath the polar cap that cannot be transferred back to the surface during the spark
development time scale (see Appendix~B in GM02). The heat flow coefficient $\kappa$ influenced the required surface
magnetic field $B_{\rm s}$ in proportion to $\kappa^{0.57}$ (for CR case). GM02 argued that for $T_{\rm s}\sim 1$~MK
this effect could reduce the required $B_{\rm s}$ by a factor of about 2. However, if $T_{\rm s}\sim 3$~MK as indicated
by the case of PSR B0943$+$10, the reduction effect is negligible and we ignored it in this paper.

\section{Discussion and Conclusions}

The phenomenon of drifting subpulses has been widely regarded as a powerful tool for understanding the mechanism of
coherent pulsar radio emission. RS75 first proposed that drifting subpulses are related to \EB\ drifting sparks
discharging the high potential drop within the inner acceleration region above the polar cap. The subpulse-associated
streams of secondary electron-positron plasma created by sparks were penetrated by much faster primary beam. This
system was supposed to undergo a two-stream instability, which should lead to generation of the coherent radiation at
radio wave lengths. However, careful calculations of the binding energy (critical for spark ignition) and the growth
rate of the two-stream instability have shown that neither the sparking discharge nor the two-stream instability were
able to work in a way proposed by RS75. Nonetheless, qualitatively their idea was still considered attractive, at least
to the authors of this paper who have been continuing efforts to search for mechanisms that would actually make the
RS75 model to work. In this paper we applied the partially screened gap model proposed by GMG03 to PSR B0943+10, a
famous drifter for which ZSP06 successfully attempted to measure the thermal X-ray from hot polar cap. We derived a
simple relationship between the X-ray luminosity $L_{\rm x}$ from the polar cap heated by sparks and the tertiary
periodicity $\hat{P}_3$  of the spark-associated subpulse drift observed in radio band. This relationship reflects the
fact that both the drifting (radio) and the heating (X-rays) rates are determined by the same value of the electric
field in the partially screened gap. As a consequence of this coupling equations~(\ref{Lx}) and (\ref{LxE}) are
independent of details of the acceleration region. In PSRs B0943$+$10 and B1133+16, the only two pulsars for which both
$L_{\rm x}$ and $\hat{P}_3$ are measured, the predicted relationship between observational quantities holds very well.
We note that $\hat{P}_3$ is very difficult to measure and its value is known only for four pulsars: B0943$+$10,
B1133+16, B0826$-$34 and B0834$+$06. The successful confrontation of the predicted X-ray luminosity with the
observations in PSRs B0943$+$10 and B1133+16 encourages further tests of the model with future X-ray observations of
other drifting pulsars. This is particularly relevant to PSR B0834$+$06, whose predicted X-ray luminosity is much
higher than in PSR B0943+10, while the distance to both pulsars is almost identical. It is worth mentioning that due to
relatively poor photon statistics it is still not absolutely clear whether the X-ray radiation associated with polar
cap of PSR B0943+10 is thermal or magnetospheric (or both) in origin. However, the case of Geminga pulsar B0633+17
strongly supports thermal origin. Observations of PSR B0834$+$06 with {\it XMM-Newton} should help to resolve this
question ultimately.

In both the steady SCLF model and the pure vacuum gap model, the potential increases with height quadratically at lower
altitudes. However the growth rate is significantly different - the latter is faster by a factor of $R/r_{\rm p}$. It
is well known that in the pure vacuum case (RS75), the potential grows so fast that a primary particle could quickly
generate pairs with a high multiplicity, and that some backward returning electrons generate more pairs and soon a
``pair avalanche'' occurs and the potential is short out by a spark. In the PSG model we are advocating, the potential
drops by a factor of $\eta$. For essentially all the cases we are discussing, this $\eta$ value is much larger than the
$r_{\rm p}/R$ value required for the steady SCLF to operate, although it is much less than unity. The steady state
condition is not satisfied, and the the gap is more analogous to a vacuum gap, i.e. the pair discharging happening
intermittently. In fact, the partially screened potential drop is still above the threshold for the magnetic pair
production, which in strong and curved surface magnetic field is a condition necessary and sufficient for the sparking
breakdown.

The original RS75 pure VG model predicts much too high a subpulse drift rate and an X-ray luminosity to explain the
case of PSR B0943+10 and other similar cases. Other available acceleration models predict too low a luminosity and the
explanation of drifting subpulse phenomenon is generally not clear at all (see ZSP05 for more detailed discussion). In
summary, the bolometric X-ray luminosity for the space charge limited flow \citep{as79} pure vacuum gap (RS75) and
partially screened gap (GMG03) is $(10^{-4} - 10^{-5})\dot{E}$ \citep{hm02}, $(10^{-1} - 10^{-2})\dot{E}$ (ZSP05), and
$\sim 10^{-3}\dot{E}$ (this paper), respectively. The latter model also predicts right \EB\ plasma drift rate. Thus,
combined radio and X-ray data are consistent only with the partially screened VG model, which requires very strong
(generally non-dipolar) surface magnetic fields. Observations of the hot-spot thermal radiation almost always indicate
bolometric polar cap radius much smaller than the canonical Goldreich-Julian value. Most probably such a significant
reduction of the polar cap size is caused by the flux conservation of the non-dipolar surface magnetic fields
connecting with the open dipolar magnetic field lines at distances much larger than the neutron star radius.

Our analysis suggests the following pulsar picture: In the strong magnetic fields approaching $10^{14}$~G at the
neutron star surface, the binding energy is high enough to prevent a full thermionic flow from the hot polar cap at the
corotation limited level. A partially screened vacuum gap develops with the acceleration potential drop exceeding the
threshold for the magnetic pair formation. The growth of this potential drop should be naturally limited by a number of
isolated electron-positron spark discharges. As a consequence, the polar cap surface is heated by back-flowing
particles to temperatures $T_{\rm s}\sim 10^6$~K, just below the critical temperature $T_{\rm c}$ at which the
thermionic flow screens the gap completely. The typical radii of curvature of the field lines ${\cal R}$ is of the
order of polar cap radii $r_{\rm p}\sim 10^3-10^4$~cm. The only parameter that is thermostatically adjusted in a given
pulsar is the shielding parameter $\eta=10^{-3}(B_{\rm s}/B_{\rm q}){\cal R}^{-0.5}_6 P \sim 0.001 T_6^{1.43}{\cal
R}^{-0.5}_6 P$, which determines the actual level of charge depletion with respect to the pure vacuum case ($\eta$=1),
and in consequence the polar cap heating rate as well as the spark drifting rate. It is worth to emphasize that
$\eta\sim 0.1$ for longer period pulsars $P\sim 1$ s and $\eta\sim 0.01$ for shorter period pulsars. Our calculations
are consistent with PSR B0943$+$10 and few other drifting pulsars, for which the signatures of X-ray emission from the
hot polar cap were detected.

The sparks operating at the polar cap generate streams of secondary electron-positron plasma flowing through the
magnetosphere. These streams are likely to generate beams of coherent radio emission that can be observed in the form
of subpulses. \citet{u87} first pointed out that the non-stationarity associated with sparking discharges naturally
leads to a two-stream instability as the result of mutual penetration between the slower and the faster plasma
components. \citet{am98} developed this idea further, calculated the growth rates, and demonstrated that the
instability is very efficient in generating Langmuir plasma waves at distances of many stellar radii $r_{\rm ins} \sim
10^{4-5} h^{CR}=10^{7-8}$ cm, where $ h^{CR}$ is the height of the acceleration region described by equation (21)
\citep{mgp00}. This is exactly where pulsar's radio emission is supposed to originate \citep[e.g.][]{kg98}. Conversion
of these waves into coherent electromagnetic radiation escaping from the pulsar magnetosphere was considered and
discussed by \citet{mgp00} and \citet{glm04}. These authors demonstrated that the nonlinear evolution of Langmuir
oscillations developing in pulsar's magnetosphere leads to formation of charged, relativistic solitons, able to emit
coherent curvature radiation at radio wavelengths. The component of this radiation that is polarized perpendicularly to
the planes of dipolar magnetic field can escape from the magnetosphere \citep[see][for observational evidence of such
polarization of pulsar radiation]{lea01}. The observed pulsar radiation in this picture is an indirect consequence of
sparking discharges within the inner acceleration region just above the polar cap. In light of this paper we can
therefore argue that the coherent pulsar radio emission is conditional on the presence of strong non-dipolar surface
magnetic fields at the polar cap, with a strength about $10^{13-14}$~G and radius of curvature of the order
$10^{4}$~cm.

In the very strong surface magnetic field assumed within the accelerator, processes such as photon splitting
\citep{bh01} and bound pair creation \citep{um95} may become important. It has been suggested that such processes could
potentially delay pair creation and thus increase the height and voltage of the accelerator. For the photon splitting
case, the delay is significant only if photons with both polarization modes split - a hypothesis in strong magnetic
fields \citep{bh01}. It could be possible that only one mode split \citep{u02}. In such a case the gap height and
potential of a PSG would not be significantly affected due to the high pair multiplicity in strong, curved magnetic
fields. For bound pairs \citep{um95,um96}, in the very hot environment near the neutron star surface (with temperatures
exceeding 1 MK), it is possible that bound pairs could not survive long from photoionization. Following \citet{bcp92}
and \citet{um96} one can roughly estimate the mean free path for bound pair photo-ionization. It turns out that for
temperatures around (2-3) MK this mean free path is of the order of few meters, which is considerably less than the
height of CR-driven PSG. In such a case, even if the bound pairs are initially produced, they would not significantly
delay the pair formation. One could address this potential potential problem by referring to the detailed case study of
PSR B0943+10 analyzed in this paper. This is the only pulsar in which we have full information concerning \EB\ drift
rate and the polar cap heating rate. In the analysis we have used two methods. The first method is independent on
details of accelerating region (such as height, potential drop, etc.) but is based only on the subpulse drift
radio-data (eqs.~[\ref{Lx},\ref{LxE}]). The second one includes the detailed treatments of model parameters without
considering the delaying effect by photon splitting and bound pairs (eqs.~[\ref{Ts},\ref{LxE2}]). Both methods give
consistent results, as illustrated in Figures (1-4). We therefore conclude that the delaying processes, if any, are not
significant in the PSGs given the physical conditions we invoke. Finally, we note that \citet{um95} presented a model
of the inner accelerator that shares some similar features with our model, although it is basically different. In their
model the potential drop is self-regulated by partial screening close to the threshold for field ionization of bound
pairs, so the surface temperature is maintained at the level necessary for this screening. In our model, the partial
screening keeps the surface temperature slightly below the critical temperature at which the thermionic flow is space
charge limited \citet{as79}. In the light of results presented in this paper we claim that bound pairs do not affect
the formation of inner accelerators in pulsars.

\begin{acknowledgements}
We acknowledge the support of the Polish State Committee for scientific research under Grant 2 P03D 029 26 (JG \& GM)
and NASA grants NNG05GB67G, NNG05GH91G and NNG05GH92G (BZ).
\end{acknowledgements}

{}

\clearpage

\begin{figure}
\plotone{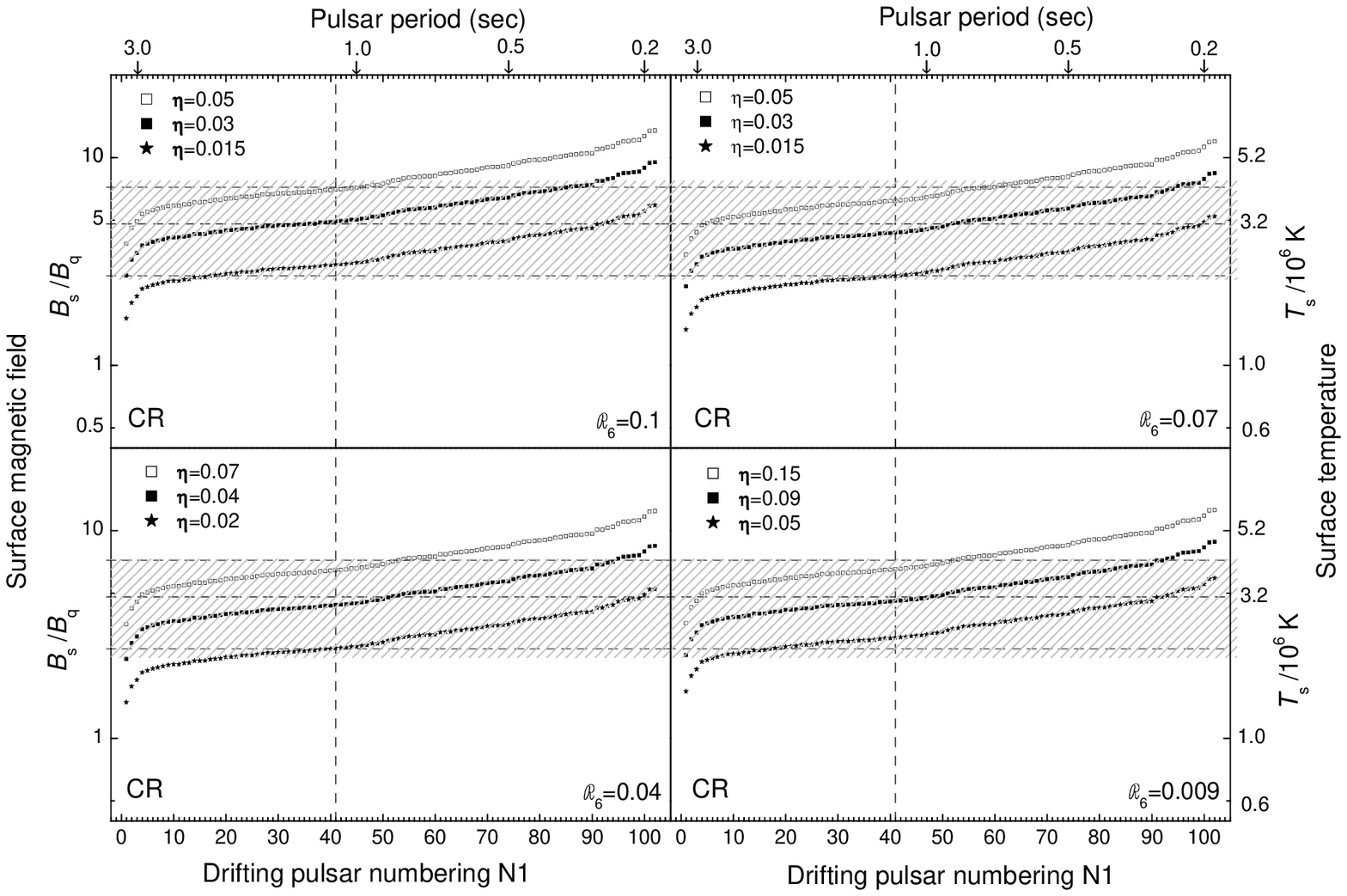}
\caption{Models of partially screened inner acceleration regions (PSG) driven by curvature radiation
seed photons above positively charged (ion case) polar cap for 102 pulsars from Table~1, sorted according to pulsar
period. The horizontal axes correspond to the pulsar number N1 (bottom) or pulsar period (top). The vertical axes
correspond to the surface magnetic field $B_{\rm s}/B_{\rm q}$ (left-hand side), or surface temperature $T_{\rm
s}/10^6$~K (right-hand side). The calculations were made for conditions corresponding to very strong ($B_{\rm
s}>5\times 10^{12}$~G) and curved ($0.1>{\cal R}_6>0.005$) surface magnetic field. The vertical lines correspond to the
case of PSR B0943$+$10 (N1=41) and the hatched area encompassing three horizontal lines correspond to the range of
surface magnetic field and temperature inferred for this pulsar from the XMM-Newton observation (ZSP05). These models
allow to read off the physical conditions existing in the acceleration region above the polar cap of a particular
pulsar in the form of parameters such as $B_{\rm s}$, ${\cal R}_6$ and $\eta$.} \label{fig1}
\end{figure}

\clearpage

\begin{figure}
\epsscale{0.70}
\plotone {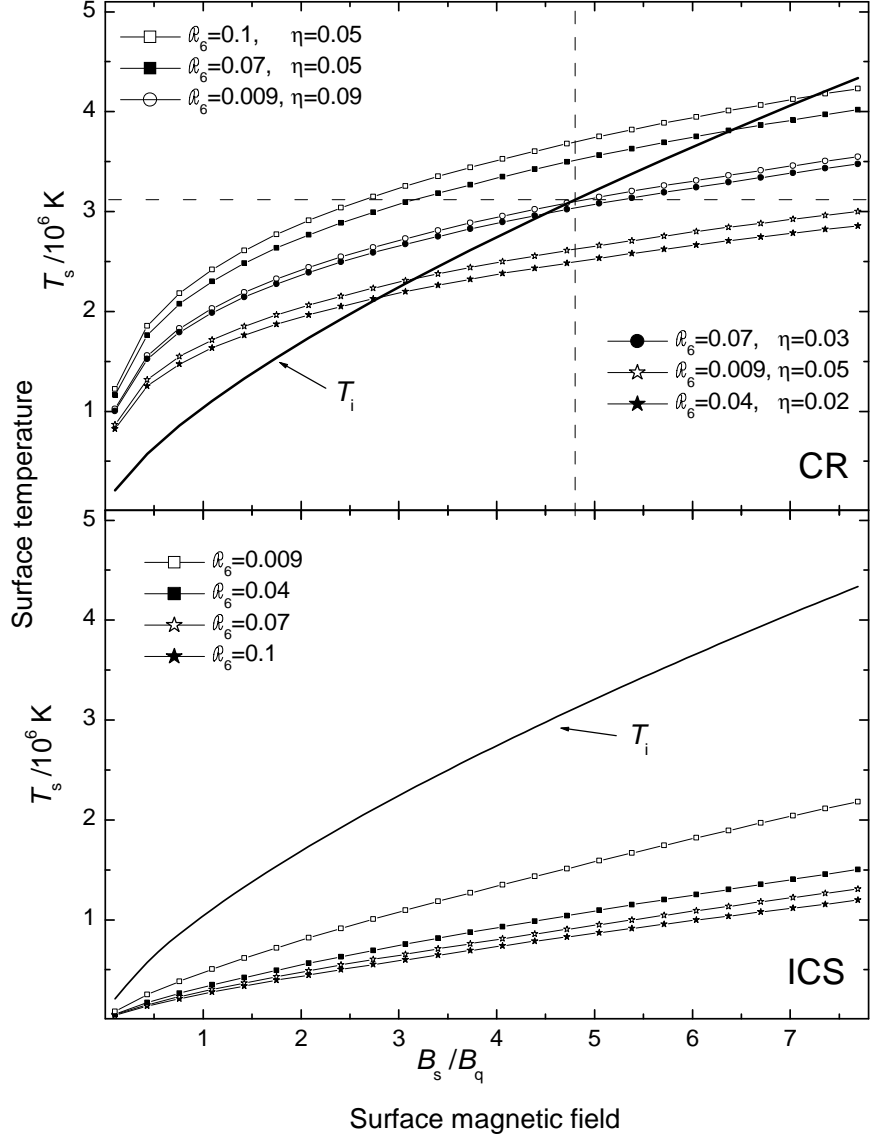} \caption{Dependence of the surface temperature on the surface magnetic field in the
case of NTPSG for CR (upper panel) and ICS (lower panel) cases. Solid lines represent the critical ion temperature
$T_{\rm i}$, while the symbol-broken lines represent the actual surface temperature $T_{\rm s}$ for different
combinations of the parameters ${\cal R}_6$ and $\eta$. As one can see only the CR case can work in pulsars with
positively charged (ion case) polar cap. The solution for PSR B0943$+$10 marked by the dashed lines is ${\cal
R}_6=0.009$, $\eta=0.09$ and $T_{\rm s}=3.1$~MK, implying $B_{\rm s}=4.8B_{\rm q}\sim 2\times 10^{14}$~G.} \label{fig2}
\end{figure}

\clearpage

\begin{figure}
\epsscale{1.0}
\plotone {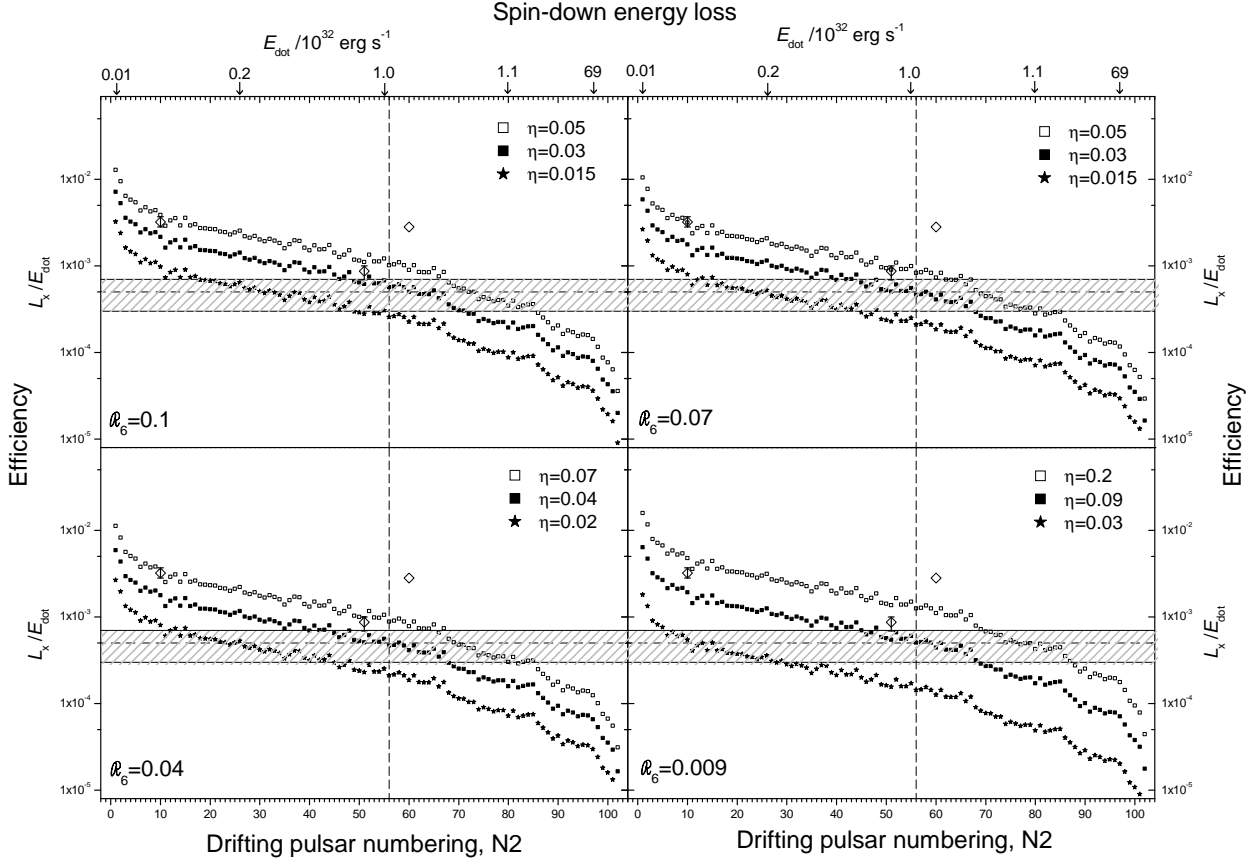}
\caption{The thermal radiation efficiency from the hot polar cap for 102 pulsars from
Table~1 for different parameters of the PSG marked in the legends. Pulsars are sorted according to the pulsar spin-down
luminosity (N2 in Table~1). For clarity of presentation the surface field is fixed at $B_{\rm s}/B_{\rm q}$=4.8. The
vertical dashed lines (N2=56) and three horizontal dashed lines ($L_{\rm x}=5 \pm 2 \times 10^{28}$ erg/s) within the
hatched area correspond to the special case of PSR B0943+10 (Tables 1 and 2). Three other pulsars for which values of
$\hat {P}_3/P$ is directly determined from single pulse radio data are included as diamond symbols: B0826-34 (N2=10),
B1133+16 (N2=51) and B0834+06 (N2=60). } \label{fig3}
\end{figure}

\clearpage

\begin{figure}
\plotone {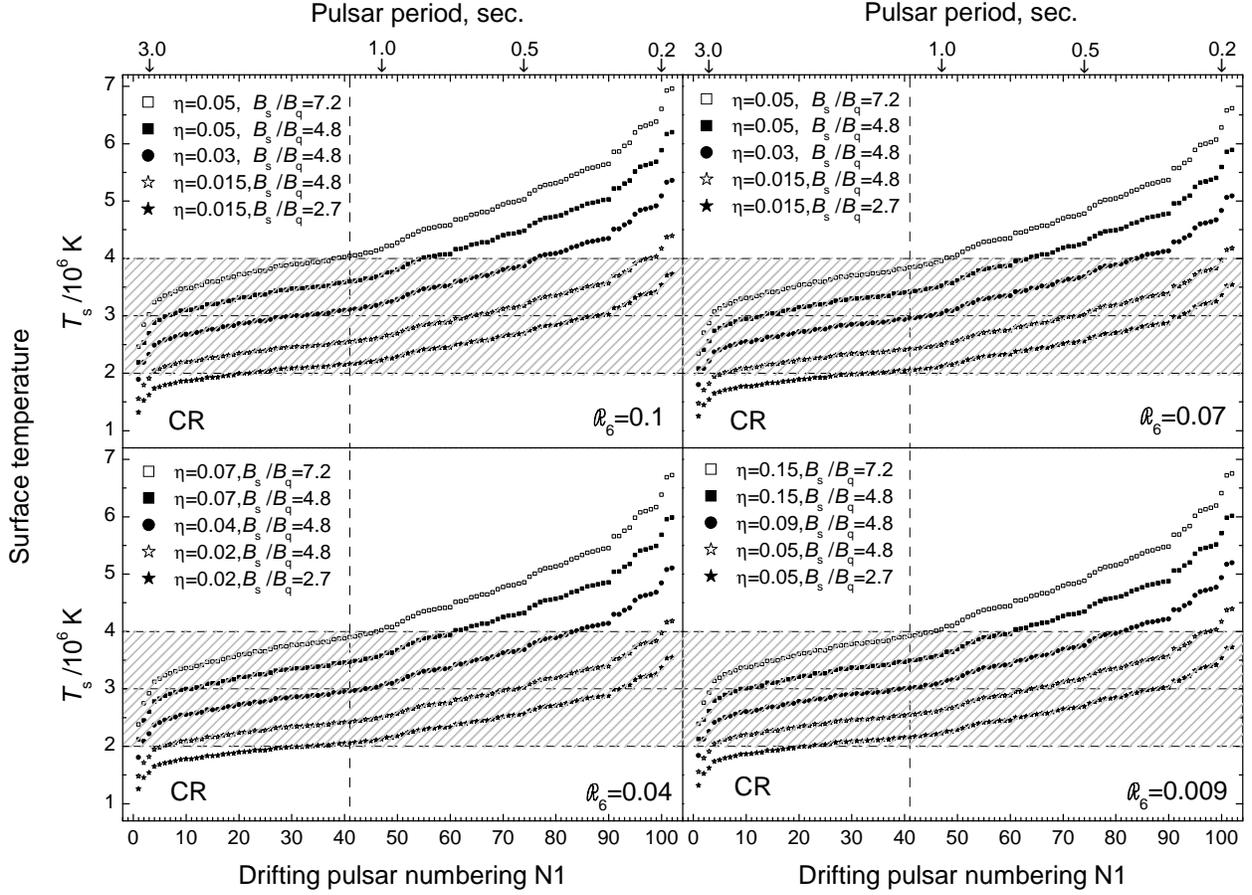}
\caption{Surface temperature of hot polar cap for 102 pulsars from Table~1 for different parameters
of the PSG marked in the legends. Pulsars are sorted according to the pulsar period (N1 in Table~1). The vertical
dashed lines (N1=41) and three horizontal dashed lines ($T_{\rm s}=3 \pm 1 \times 10^6$ K) within the hatched area
correspond to the case of PSR B0943+10. }
\label{fig4}
\end{figure}

\clearpage

\begin{table}
\begin{center}

\fontsize{8}{8pt}\selectfont

\caption{Comparison of observed and predicted parameters}

\begin{tabular}{c c c c c c c c c c c c}

\tableline
\tableline
Name&\multicolumn{2}{c}{$\hat{P}_{3}/P$}&\multicolumn{2}{c}{${L_{\mathrm{x}}}/{\dot{E}}\times 10^{-3}$}&
\multicolumn{2}{c}{$L_{\mathrm{x }}\times 10^{28}$} & $b$ & $T_{\rm s}^{\mathrm{(obs)}}$ &
$T_{\rm s}^{\mathrm{(pred)} }$ & $B_{\mathrm{d}}$ & $B_{\mathrm{s}}$ \\
\ PSR B & Obs. & Pred. & Obs. & Pred. & Obs. & Pred. &
${A_{\mathrm{pc}}}/{A_{\mathrm{bol}}}$ & $10^{6}\ $K & $10^{6}\ $K & $10^{12}$G & $10^{14}$G \\
\tableline
$0943+10$&$37.4$&$36$&$0.49_{-0.16}^{+0.06}$&$0.45$&$5.1_{-1.7}^{+0.6}$&$4.7$&$60_{-48}^{+140}$&$3.1_{-1.1}^{+0.9}$&$3.3_{-1.1}^{+1.2}$&$3.95$&$2.37_{-1.90}^{+5.53}$\\
$1133+16$&($33_{-3}^{+3}$)&$27_{-2}^{+5}$&$0.77_{-0.15}^{+0.13}$&$0.58_{-0.09}^{+0.12}$&$6.8_{-1.3}^{+1.1}$&$5.1_{-0.8}^{+1.0}$&$11.1_{-5.6}^{+16.6}$&$2.8_{-1.2}^{+1.2}$&$2.1_{-0.4}^{+0.5}$&$4.25$&$0.47_{-0.24}^{+0.71}$ \\
$0826-34$ & $14_{-1}^{+1}$ &  &  & $3.2_{-0.4}^{+0.5}$ &  & $ 2.0_{-0.25}^{+0.33}$ &  &  &  & $2.74$ &  \\
$0834+06$ & $15$ &  &  & $2.8$ &  & $37$ &  &  &  & $5.94 $ &  \\
$0809+74$ & $150$ &  &  & $0.003$ &  & $0.009$ & &  &  & $0.5$   &  \\
$0633+17$ & & $400$ & $0.004$ & & $15$ & & $25$ & $1.9_{-0.3}^{+0.3}$ & $2.13$ & $1.6$ & $0.4$ \\
\tableline
\end{tabular}
\end{center}
\end{table}

\clearpage

\begin{table}
\caption{List of 102 pulsars with drifting subpulses compiled from
R86, GM02 and WES05. N1 and N2 refer to sorting according to
pulsar period $P$ and spin-down luminosity $\dot{E}$, respectively.}
\label{Table}
\begin{center}
{\footnotesize \begin{tabular}{lllllllllllll} \tableline\tableline
Name      &N1  &N2 &Name       &N1  &N2 &Name       & N1 &N2 &Name       & N1 &N2\\
\tableline
B0011+47  &31 &17 &B0823+26   &71 &71 &B1822-09   &54 &91 &B1953+50   &73 &69\\
B0031-07  &49 &25 &B0826-34   &10 &10 &J1830-1135 &1  &11 &B2000+40   &50 &53\\
B0037+56  &40 &50 &B0834+06   &28 &60 &B1839+56   &15 &18 &J2007+0912 &76 &61\\
B0052+51  &6  &40 &B0919+06   &78 &97 &B1839-04   &11 &5  &B2011+38   &97 &100\\
B0136+57  &94 &99 &B0940+16   &42 &3  &B1841-04   &47 &64 &B2016+28   &69 &37\\
B0138+59  &33 &13 &B0943+10   &41 &56 &B1844-04   &67 &98 &B2020+28   &89 &86\\
B0144+59  &100&85 &B1039-19   &24 &19 &B1845-01   &61 &76 &B2021+51   &72 &79\\
B0148-06  &20 &8  &B1112+50   &14 &28 &B1846-06   &21 &74 &B2043-04   &18 &22\\
B0301+19  &23 &24 &B1133+16   &36 &51 &B1857-26   &65 &38 &B2044+15   &38 &7\\
B0320+39  &3  &2  &B1237+25   &25 &20 &B1859+03   &62 &80 &B2045-16   &8  &46\\
B0329+54  &59 &66 &B1508+55   &56 &72 &B1900+01   &57 &70 &B2053+36   &99 &84\\
B0450+55  &90 &87 &B1530+27   &39 &29 &J1901-0906 &12 &16 &B2106+44   &81 &44\\
B0523+11  &87 &48 &B1540-06   &60 &54 &B1911-04   &52 &68 &B2110+27   &35 &47\\
B0525+21  &2  &35 &B1541+09   &55 &41 &B1914+13   &93 &96 &B2111+46   &46 &32\\
B0540+23  &95 &101&B1604-00   &80 &65 &J1916+0748 &70 &88 &B2148+63   &84 &58\\
B0609+37  &92 &52 &B1612+07   &34 &45 &B1917+00   &29 &63 &B2154+40   &19 &39\\
B0621-04  &45 &33 &B1642-03   &83 &83 &B1919+21   &26 &27 &B2255+58   &85 &92\\
B0626+24  &75 &75 &J1650-1654 &13 &30 &B1923+04   &43 &49 &B2303+30   &17 &34\\
B0628-28  &30 &62 &B1702-19   &91 &95 &B1924+16   &68 &89 &B2310+42   &88 &55\\
B0727-18  &74 &94 &B1717-29   &64 &59 &B1929+10   &98 &90 &B2319+60   &5  &31\\
B0740-28  &101&102&B1737+13   &53 &57 &B1933+16   &86 &93 &B2324+60   &96 &81\\
B0751+32  &22 &21 &B1738-08   &7  &14 &B1937-26   &82 &73 &B2327-20   &16 &42\\
B0809+74  &27 &4  &B1753+52   &4  &6  &B1942-00   &44 &23 &J2346-0609 &37 &36\\
J0815+0939&63 &26 &B1804-08   &102&67 &B1944+17   &77 &15 &B2351+61   &48 &77\\
B0818-13  &32 &43 &B1818-04   &66 &82 &B1946+35   &58 &78 &   &   &\\
B0820+02  &51 &9  &B1819-22   &9  &12 &B1952+29   &79 &1  &   &   &\\
\tableline
\end{tabular}}
\end{center}
\end{table}

\label{lastpage}

\end{document}